\documentclass[fleqn,usenatbib]{mnras}

\usepackage{newtxtext,newtxmath}
\usepackage{siunitx} 
\usepackage{listings}
\usepackage{hyperref}
\usepackage{booktabs}
\usepackage{capt-of}
\usepackage{upgreek}
\usepackage{caption}
\usepackage{footmisc}
\usepackage{academicons}
\usepackage{xcolor}
\usepackage{subcaption}
\usepackage[authoryear]{natbib}
\usepackage{lipsum}  
\usepackage{xfrac}

\newcommand{\milleniumII}{\texttt{Millenium-II}}

\usepackage[T1]{fontenc}

\DeclareRobustCommand{\VAN}[3]{#2}
\let\VANthebibliography\thebibliography
\def\thebibliography{\DeclareRobustCommand{\VAN}[3]{##3}\VANthebibliography}

\usepackage{graphicx}	
\usepackage{amsmath}	
\usepackage{multirow}
\usepackage{float}
\usepackage{color}
\definecolor{myorange}{rgb}{0.8, 0.3, 0.0}

\newcommand{\orcid}[1]{\href{https://orcid.org/#1}{\textcolor[HTML]{A6CE39}{\includegraphics[width=8pt]{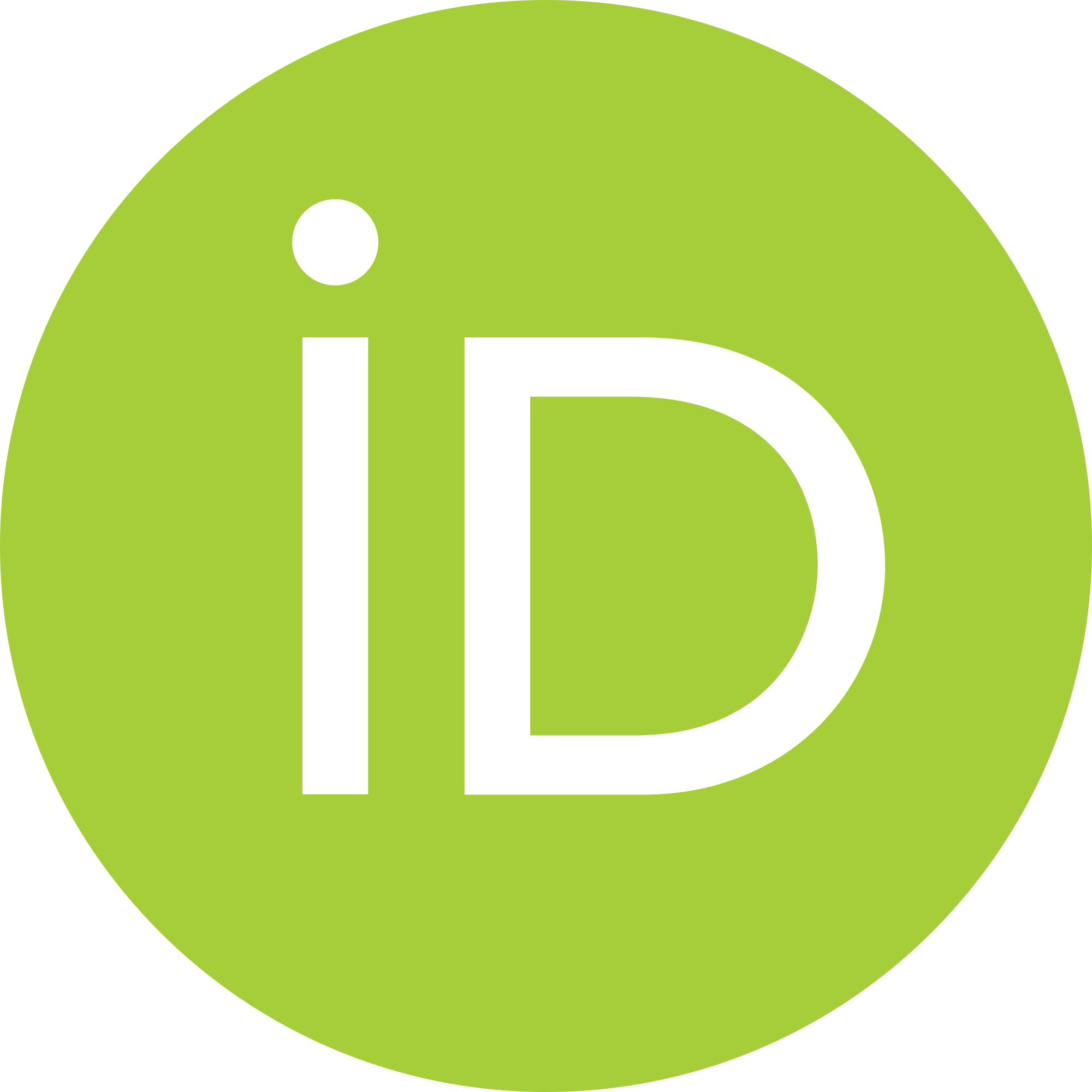}}}}

\title[]{A link to the past: characterizing wandering black holes in Milky Way-type galaxies}

\author[J. Untzaga et al.]{
J.Untzaga$^{1,2}$ \orcid{0000-0001-9140-8113},
S. Bonoli$^{2,3}$ \orcid{0000-0002-6381-2052},
D. Izquierdo-Villalba$^{4,5}$ \orcid{0000-0002-6143-1491},
M. Mezcua$^{1,6}$ \orcid{0000-0003-4440-259X}
D. Spinoso$^{7}$ \orcid{0000-0002-9074-4833,
} 
\thanks{E-mail: juntzaga96@gmail.com} 
\\
$^{1}$Institute of Space Sciences (ICE, CSIC), Campus UAB, Carrer de Magrans, 08193 Barcelona, Spain\\
$^{2}$Donostia International Physics Center (DIPC), Manuel Lardizabal Ibilbidea, 4, San Sebastián, Spain\\
$^{3}$IKERBASQUE, Basque Foundation for Science, E-48013, Bilbao, Spain \\
$^{4}$Dipartimento di Fisica “G. Occhialini”, Università degli Studi di Milano-Bicocca, Piazza della Scienza 3, I-20126 Milano, Italy\\
$^{5}$INFN, Sezione di Milano-Bicocca, Piazza della Scienza 3, 20126 Milano, Italy\\
$^{6}$Institut d’Estudis Espacials de Catalunya (IEEC), Edifici RDIT, Campus UPC, 08860 Castelldefels (Barcelona), Spain\\
$^{7}$Department of Astronomy, 6th floor, MongManWai Building, Tsinghua University, Beijing 100084, People's Republic of China
}

\date{}

\pubyear{2023}

\begin{document}
\label{firstpage}
\pagerange{\pageref{firstpage}--\pageref{lastpage}}
\maketitle

\begin{abstract}
A population of non-stellar black holes ($\gtrsim$100 M$_{\odot}$) has been long predicted to wander the Milky Way. We aim to characterize this population by using the \texttt{L-Galaxies} semi-analytical model applied on top of the high resolution \texttt{Millennium-II} merger trees. Our results predict $\sim$10 wandering black holes with masses $\sim$2 $\times$ 10$^{3}$ M$_{\odot}$ in a typical $z$ = 0 Milky Way galaxy, accounting for $\sim$2$\%$ of the total non-stellar black hole mass budget of the galaxy. We find that the locations of these wanderers correlate with their formation scenario. While the ones concentrated at $\lesssim$1 kpc from the galactic nucleus on the disc come from past galactic mergers, the ones formed as a consequence of ejections due to gravitational recoils or the disruption of satellite galaxies are typically located at $\gtrsim$100 kpc. Such small and large distances might explain the absence of strong observational evidence for wandering black holes in the Milky Way. Our results also indicate that $\sim$67$\%$ of the wandering population is conformed by the leftovers of black hole seeds that had little to no growth since their formation. We find that wandering black holes that are leftover seeds become wanderers at an earlier time with respect to grown seeds, and also come from more metal-poor galaxies. Finally, we show that the number of wandering black holes in a Milky Way-type galaxy depends on the seeding efficiency.


\end{abstract}

\begin{keywords}
methods: numerical -- software: simulations -- Galaxy: evolution -- black hole physics
\end{keywords}



\section{Introduction}
\label{Introduction}
While the formation mechanism of stellar-mass black holes (BHs) is well established, the origin of supermassive BHs ($\geq$10$^{6}$ M$_{\odot}$) is unclear (see, e.g.,  \citealt{inayoshi2020, volonteri2021} for a full review). Observational evidence of more than 300 supermassive BHs at z $\sim$ 7 \citep{fan2001, fan2003, willott2007, willott2010, mortlock2011, venemans2013, wu2015, bañados2018, decarli2018, matsuoka2019, wang2021, naidu2022, castellano2022, adams2023} and even z $\sim$ 11 \citep{maiolino2024} requires the formation of BH  ``seeds'' at even higher redshift from which these supermassive BHs grew via accretion and mergers. 

Possible seed BH formation channels include (1) the direct collapse of low-metallicity Population III stars of a few hundred solar masses \citep{fryer2001, madau2001, heger2002, heger2003, spera2017}, (2) the  direct collapse of rapidly inflowing dense gas in the first protogalaxies \citep{loeb1994, eisenstein1994, bromm2003, lodato2006}, (3) the direct collapse induced by the major merger of massive gas-rich galaxies \citep{mayer2018} and (4) runaway collisions of stars in dense stellar clusters \citep{coleman2002, oleary2006, devecchi2009, giersz2015, mapelli2016}. Theoretical models support that a fraction of these seeds did not grow into supermassive BHs, and instead remained with a similar mass up until current redshift (e.g., \citealt{greene2012, reines2016, spinoso2023}). This means that characterizing the properties and formation mechanisms of leftover seeds found in the local Universe is crucial for a thoroughgoing understanding of the non-stellar BH population all along its history. But where can we find these seeds?

One of the potential locations for leftover seeds could be the core of dwarf galaxies, where seed BHs might not have had the chance to grow substantially \citep{volonteri2010, greene2012, reines2016}. Indeed, black holes in the intermediate mass regime (10$^{2}$~M$_{\odot}$ -- 10$^{6}$ M$_{\odot}$, IMBH onwards) could be leftover seeds. The first observational evidence for IMBHs traces back to the late 80s, with the elliptical galaxy Pox 52 \citep{kunth1987} and the spiral galaxy NGC 4395 \citep{filippenko1989}. To detect BHs through kinematics, one must be able to resolve the stellar motions that take place within their gravitational sphere of influence. Since this type of observation is greatly dependant on the spatial resolution of the sample, the limiting distance for kinematic observations is set to $\sim$4 Mpc \citep{nguyen2018, greene2020}. For this reason, a variety of methods can be employed in the absence of kinematic signatures or to complement them. For example, IMBHs in dwarf galaxies are often found in active galactic nuclei (AGNs) through optical spectroscopy \citep{greene2004, greene2007, dong2007, dong2012, izotov2007, reines2013, chilingarian2018, mezcua2020broad, salehirad2022, mezcua2024}, through the detection of hard X-ray emission or radio emission \citep{reines2011, reines2014, reines2020, mezcua2016, mezcua2018, mezcua2019z3.4} or even by utilizing near-infrared photometry \citep{fischer2006, giallongo2015, bohn2021}.

IMBHs can also be found as the result of in-situ formation in nearby globular clusters \citep{kiziltan2017}. The presence of seeds in globular clusters was first suggested by \citet{silk1975} when studying their X-ray emission. Since then, many studies have aimed at detecting the radiative accretion signature of IMBHs within globular clusters, but no conclusive results have been obtained (e.g., \citealt{maccarone2005, bash2007, maccarone2008, cseh2010, strader2012, haggard2013, wrobel2015, mezcua2017, tremou2018, baumgardt2019, wrobel2020}). While several candidates have been found from dynamical models, there has been no direct X-ray nor radio detections of the sources. A few minor exceptions arise, like globular cluster G1 in M31 \citep{pooley2006, kong2007, ulvestad2007} and more recently the outbursting source 3XMM J215022.4-055108 \citep{lin2018, lin2020}, both of them detected in the X-ray regime. In any case, since globular clusters have scarce amounts of gas and dust, any signatures of accretion from a putative leftover seed in the X-ray and radio regimes are expected to be very low. 

There is also observational evidence of non-nuclear IMBHs in the form of of ultraluminous and hyperluminous X-ray sources (ULXs and HLXs, respectively; e.g., \citealt{matsumoto2001, farrell2009, jonker2010, mezcua2011, mezcua2013a, mezcua2013b, kim2015, mezcua2015, comerford2015, lin2016, mezcua2018, lin2020}). ULXs are extragalactic and off-nuclear X-ray sources with luminosities L$_{\mathrm{X}}$~$\geq$~10$^{39}$ erg/s that are less luminous than AGNs but more luminous than any other stellar process with intrinsic X-ray luminosities. Since most of this L$_{\mathrm{X}}$ $\sim$ 10$^{39}$ erg/s sources can be explained by stellar-mass BHs undergoing super-Eddington accretion, HLXs, characterized by their even higher X-ray luminosities (L$_{\mathrm{X}}$ $\geq$ 10$^{41}$ erg/s) remain the best seed BH candidates \citep{barrows2019}. Some examples include HLX-1 (e.g., \citealp{farrell2009, davis2011, webb2012}), M82-X1 \citep{kaaret2001, pasham2014} and NGC 2276-3c \citep{sutton2012, mezcua2013b, mezcua2015}, three ULXs that would have belonged to the nucleus of a dwarf galaxy before being stripped during a minor merger with the current host galaxy \citep{king2005, soria2013, mezcua2015}. 

Plenty of observational evidence has been found regarding IMBHs, both central and non-nuclear, both at local Universe and up to redshift z $\sim$ 3.4 \citep{mezcua2019z3.4}; and since most of these detections are carried out by examining radiative signatures, it is safe to assume that an even greater population of IMBHs is yet to be found when also accounting for non-active ones. This population --a fraction of which is theorized to correspond to leftover seeds-- would have been accreted via hierarchical merging into the Milky Way (MW) throughout its history \citep{volonteri2005, bellovary2010}. At $z$~=~0, these leftover seeds should be wandering within the MW, thus contributing to its population of wandering black holes (WBHs).

The study of WBHs has recently gained significant attention thanks in part to the advancements in numerical simulations of galaxy formation and evolution \citep{volonteri2003, volonteri2005, bellovary2010, gonzalez2018, izquierdo2020, greene2021, ricarte2021b}. In terms of the total expected amount of WBHs in MW-type galaxies, \cite{ricarte2021} used the \texttt{ROMULUSC} zoom-in cosmological simulation on a 10$^{14}$ M$_{\odot}$ galaxy cluster to predict over one thousand WBHs and found a correlation between the expected number of wanderers in a galaxy and its halo mass. Re-scaling to the MW halo mass, $\sim$10 WBHs are predicted to wander our galaxy. This prediction is in agreement with a previous work by \cite{tremmel2018} where they studied a sample of 26 MW-mass haloes extracted from the \texttt{ROMULUS25} cosmological simulation. They found $\sim$12 wanderers within 1 virial radius of a MW-mass galaxy, $\sim$5 of them within 10 kpc of the galactic centre. In agreement with a higher concentration of WBHs in the central regions, \cite{ricarte2021} found that 50$\%$ of all wanderers are within 0.1 virial radii. Following a similar trend, \cite{weller2022} used the \texttt{illustris TNG50} simulation to study the kinematics and dynamics of star clusters acting as IMBH proxies in MW analogues, and they found that over 4$\%$ of the total population of wanderers was predicted within 1 kpc of the galactic centre. 

We find a more optimistic prediction of the overall number of WBHs in a MW analog in the work of \cite{greene2021}, who used the semi-analytic model SatGen \citep{jiang2021} to estimate $\sim$100 expected WBHs in the MW halo. However, this prediction assumes that every infalling dwarf satellite hosts a central BH that will later become a WBH of the MW-type galaxy. \cite{oleary2009,oleary2012} used $\gtrsim$1000 merger tree histories of the MW to calculate the number and mass distribution of those WBHs that may have been ejected from their small host galaxies during the MW formation history. They found that $\gtrsim$100 WBHs with masses $\gtrsim$10$^{4}$ M$_{\odot}$ should be in the MW halo today, surrounded by compact star clusters that make up for $\sim$1$\%$ of their WBH’s mass. 

\cite{kruijssen2019} compared the age-metallicity of 96 MW globular clusters to the globular cluster population of the E-MOSAICS project, consisting of 25 cosmological zoom-in simulations of MW-mass galaxies based on the EAGLE galaxy formation simulations. The study concluded that the MW has experienced $\sim$15 accretion events from $\gtrsim$4.5 $\times$ 10$^{6}$ M$_{\odot}$ satellites. Of these $\sim$15 accretion events, the MW experienced $\sim$9 mergers at $z$ > 2 and $\sim$6 mergers at $z$ < 2. On the other hand, \cite{weller2022} used the Illustris TNG50 cosmological simulation to determine that $\sim$87$\%$ of WBHs in a MW-type galaxy exhibit a tendency to drift towards the galactic centre, with sinking time-scales $\sim$10 Gyr. Combining the findings of \cite{kruijssen2019} (who observed mergers at $z$ > 2 and $z$ < 2) and \cite{weller2022} (who established an upper bound limit of sinking time-scales $z$ $\sim$ 2), it seems plausible that both a population of non-disrupted infalling satellites hosting non-stellar BHs and a population of recoiled WBHs should be expected in MW-type galaxies at the current redshift.

As for the spatial location of these wanderers, \cite{tremmel2018} estimated with a >4$\sigma$ significance that WBHs are preferentially found outside of the Galactic disc. This could incentivize a search for small, compact star clusters around these wanderers in the low-density environment of the halo. However, most of the WBHs seem not to have retained a stellar counterpart or to be located at larger fractions of the virial radius (\citealt{ricarte2021}).

The aim of this research is to characterize both the WBH and the leftover seed populations in the MW at $z$ = 0, and to improve our understanding of their merger history. We expect to (1) improve our knowledge about the co-evolution of WBHs and galaxies, specifically the final abundance and radial distribution of WBHs in MW-type galaxies, (2) obtain useful information that can help interpret past observational searches of WBHs as well as guide future ones (see \S \ref{WBH population} for more details) and (3) study the early BH seeding efficiency in the model and its impact to the final population of WBHs. This characterization has been carried out by using the semi-analytic simulation for galaxy evolution \texttt{L-Galaxies} (\citealt{henriques2015}; see \S \ref{section L-Galaxies}). In \S \ref{results} we study the populations of WBHs and leftover seeds at $z$ = 0 and the time of their formation, and we also review the effects of varying the initial seeding prescription in the model. We summarize and discuss our results in \S \ref{Conclusions good}. A Lambda Cold Dark Matter $(\Lambda$CDM) cosmology with parameters $\Omega_{\rm m} \,{=}\,0.315$, $\Omega_{\rm \Lambda}\,{=}\,0.685$, $\Omega_{\rm b}\,{=}\,0.045$, $\sigma_{8}\,{=}\,0.9$ and $\rm H_0\,{=}\,67.3\, \rm km\,s^{-1}\,Mpc^{-1}$ is adopted throughout the paper \citep{planckcollaboration2014}.

\section{The \texttt{L-Galaxies} semi-analytical model}
\label{section L-Galaxies}
In this work, we make use of the \texttt{L-Galaxies} semi-analytic code (\citealt{guo2011, henriques2015}; see \S \ref{lgalaxies}), which was designed to model galaxy formation and evolution on the merger trees of the \texttt{Millennium} (\texttt{MR},  \citealt{springel2005}) and \texttt{Millennium-II} (\texttt{MR-II}, \citealt{boylan2009}) cosmological, N-body simulations. Since we are particularly interested in studying the WBH population and its link with the leftover seeds from the high-$z$ Universe, we make use of the \texttt{MR-II} merger trees, whose mass resolution allows us to trace the formation and evolution of WBHs in dark matter (DM) haloes of 10$^{8}$ -- 10$^{14}$ M$_{\odot}$. In the following subsections, we briefly describe the properties of the \texttt{MR-II} simulation and the baryonic physics included in \texttt{L-Galaxies} to trace the formation and evolution of galaxies and their WBHs.

\subsection{The DM component: The Millenium-II simulation}
\label{milleniumII}
$\milleniumII$ is a large DM N-body simulation that follows 2160$^{3}$ particles of 6.89 $\times$ 10$^{6}$ h$^{-1}$ M$_{\odot}$ within a 100 h$^{-1}$ Mpc comoving box. Structures formed in the simulation (i.e., the so-called DM haloes) are identified and stored at 68 different epochs or \textit{snapshots} by making use of the \texttt{SUBFIND} algorithm. DM haloes hosting at least more than 20 particles are stored and arranged according to their evolutionary path in the so-called merger trees \citep{springel2005}. Taking into account the particle mass resolution and the box size, the minimum and maximum mass of a DM halo inside the \texttt{MR-II} merger trees corresponds to 1.38 $\times$ 10$^{8}$ M$_{\odot}$ and 8.22 $\times$ 10$^{14}$ M$_{\odot}$, respectively. This makes \texttt{MR-II} the perfect simulation to trace not only the hosts of MW-type galaxies but also the ones of the dwarf satellites that at some point might deposit their WBH population in them. Even though the finite time resolution offered by the \texttt{MR-II} merger trees (${\sim}\,300\,\rm Myr$) is adequate to study the evolution of DM haloes, it hinders the possibility of tracing accurately the baryonic physics involved in galaxy evolution (with dynamical times that can be up as short as few Myrs). To overcome this limitation, \texttt{L-Galaxies} does an internal time discretization between the $\milleniumII$ outputs with a time resolution that varies between $5\,{-}\,20\,\rm Myr$, depending on redshift.

\subsection{The treatment of the galaxies}
\label{lgalaxies}
\texttt{L-Galaxies} is a state-of-the-art semi-analytical model that couples many different astrophysical processes with DM merger trees extracted from N-body simulations. In this work, we use as a foundation the version presented in \cite{henriques2015} with the modifications of \cite{izquierdo2019}, \cite{izquierdo2020} and \cite{izquierdo2022}, included to improve the predictions for galaxy morphology, extend the physics of WBHs and introduce the formation and evolution of BH binaries.

The model assumes that as soon as a DM halo is formed, a fraction of the baryonic component of the universe collapses within it. This baryonic component has a mass that corresponds to 15.5$\%$ of the DM halo mass, and is given by the cosmic mean baryon fraction for the \textit{Planck} cosmology (see \citealt{henriques2015}). The baryon component is heated and distributed within the halo in the form of a diffuse, spherical and quasi-static hot gas atmosphere. The hot gas atmosphere undergoes a cooling process regulated by the functions of \cite{sutherland1993}, and, as a result, a fraction of the gas condensates and migrates towards the centre of the DM halo. The continuous cooling results in the cold gas component becoming massive enough for star formation to take place, leading to the formation of a stellar disc \citep{white1978, white1991, springel2001}. 

Right after the star formation events, massive and short-lived stars explode and inject energy into the surrounding medium via supernovae feedback, warming up and expelling part of the galaxy’s cold gas. Besides supernovae feedback, the model also invokes the so-called radio-mode feedback from central supermassive BHs, which efficiently suppresses gas cooling in massive galaxies. On top of the secular evolution, galaxies interact among themselves thanks to the hierarchical assembly of DM haloes. 

\texttt{L-Galaxies} tracks the formation and evolution of orphan galaxies at each time-step. There are two scenarios in which a galaxy can lose its DM halo and become an orphan galaxy. In the first scenario, a larger DM halo strips DM particles from a smaller satellite halo that hosts a galaxy. As a result, the satellite DM halo is left with fewer than 20 DM particles and will no longer be resolved by the simulation in the next snapshot. The galaxy from the satellite DM halo has lost its halo and is now tagged as being an orphan. In this case, the galaxy with which the orphan galaxy will merge corresponds to the one hosted in the central halo of the friend-of-friend group. In the second scenario, a DM halo hosting a galaxy is absorbed (or merged) by a central DM halo. This galaxy, usually a dwarf satellite although not necessarily, is now considered an orphan galaxy and starts infalling towards the centre of the massive DM halo. The duration of the merging process is ruled by the dynamical friction presented in \cite{binney1987} and the decay of the satellite orbit is modeled following \cite{guo2011} (see also supplemental material of \citealt{henriques2015} for further information). 

These interactions are classified by \texttt{L-Galaxies} as major and minor depending on the mass ratio of the two merging systems, and their outcomes are completely different. While major interactions are assumed to be able to destroy the disc component of central galaxies and lead to the formation of pure elliptical galaxies, minor interactions do not have an effect on the disc of central galaxies and instead trigger the build-up of a stellar bulge due to the absorption of satellite galaxies. Finally, environmental processes are also accounted for by \texttt{L-Galaxies}. Specifically, the infalling of galaxies in more massive haloes causes the loss of the entire galaxy's hot gas atmosphere and the removal of its cold gas and stellar component through tidal forces (ram pressure and galaxy tidal disruption processes, respectively). 

\subsection{The treatment of BHs and the formation of WBHs}
 \label{BH seeding}
On top of the galaxy physics, the version of \texttt{L-Galaxies} used in this work includes a sophisticated model to treat the formation and evolution of BHs, the dynamical evolution of BH binaries and the formation and evolution of WBHs. In the following section, we outline the relevant physics for this work.

\subsubsection{Formation and growth of BHs} \label{formation and growth}

\begin{figure}
    \begin{center}
    \includegraphics[scale = 0.6]{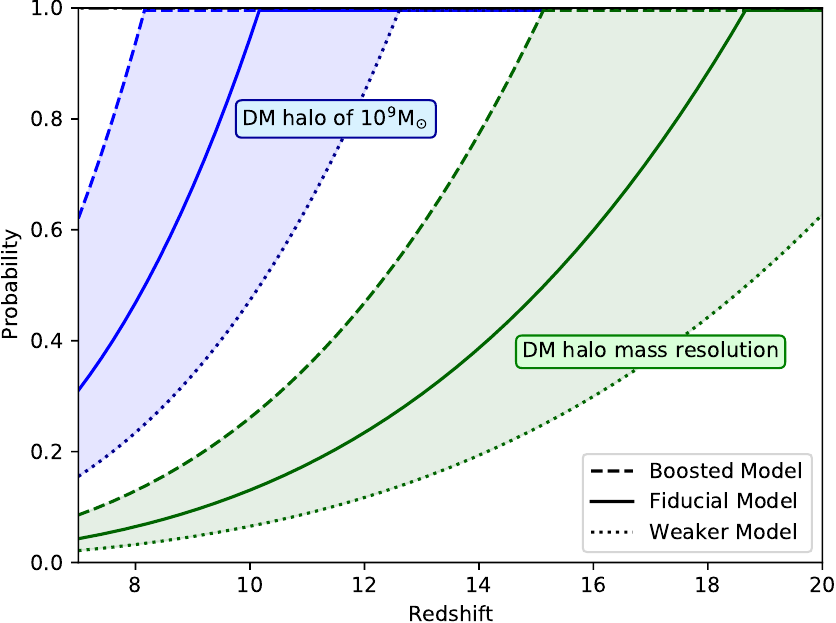} 
     \caption{Probability used by \texttt{L-Galaxies} to seed a central BH in function of redshift in a DM halo of 10$^{9}$ M$_{\odot}$ (blue) and in a DM halo with the minimum mass allowed of 1.38 $\times$ 10$^{8}$ M$_{\odot}$ (green). The solid lines indicate the probability function followed by the standard seeding prescription (see Eq.\ref{probability equation}). The dashed lines represent the probability function of the boosted seeding while the dotted lines indicate the probability function of the weaker seeding.}

    \label{probability}
    \end{center}
    \end{figure}

The formation of the first BHs in \texttt{L-Galaxies} has been extensively studied in \cite{spinoso2023}. Specifically, the authors included the genesis of light seeds (Population III remnants) and massive seeds (i.e. intermediate-mass and heavy BH-seeds) by using sub-grid approaches and accounting for the local spatial variations of the intergalactic metallicity and the UV-background produced by star formation events. Following this results, in this work we address the genesis of the first BHs in a much more simplified way, taking into account the analytical seeding probability presented in \cite{spinoso2023} and modified by \cite{izquierdo2023}. In brief, at any redshift, the probability that a newly resolved halo inside the \texttt{MR-II} merger trees is seeded with a BH is similar to the one predicted by the complete model of \cite{spinoso2023}. In this way, every time a halo is resolved at $z$ $\geq$ 7 (i.e. redshift at which galaxies tend to be pristine in \texttt{L-Galaxies}), it has a probability $\mathcal{P}$ of being seeded with a central BH with mass randomly chosen from [10$^{2}$ -- 10$^{4}$] M$_{\odot}$ (see Figure 3 of \citealt{spinoso2023}):
\begin{equation}
\label{probability equation}
    \mathcal{P} \ = \ \mathcal{A} \ (1 \ + \ z)^{\gamma} \ \left (  \frac{M_{\text{halo}}}{M^{\text{th}}_{\text{halo}}}\right )  
\end{equation}
\noindent where $M_{\rm halo}$ is the virial mass of the newly resolved halo, $z$ its redshift and $\mathcal{A}$, $\gamma$,  $M^{\text{th}}_{\text{halo}}$ three free parameters set to 0.015, 7/2 and 7 $\times$ 10$^{10}$~M$_{\odot}$, respectively.
To guide the reader, Fig. \ref{probability} shows how the seeding probability varies with redshift in the case of two representative DM halo masses, 10$^{9}$ M$_{\odot}$ and the DM halo mass resolution of \texttt{Millenium-II}, which corresponds to 1.38 $\times$ 10$^{8}$ M$_{\odot}$. To quantify how the uncertainties related to BH formation impact the population of WBHs in MW-type galaxies, we will assume different values for the parameter $\mathcal{A}$ in Eq.(\ref{probability equation}). Specifically, we will explore (1) a boosted seeding scenario ($\mathcal{A}$$_{\text{boosted}}$ = 2.0$\mathcal{A}$) where the galaxies have a larger probability of being seeded with respect to the fiducial model, and (2) a weaker seeding scenario ($\mathcal{A}$$_{\text{weaker}}$ = 0.5$\mathcal{A}$) where galaxies are less frequently seeded than in the fiducial case (see Fig. \ref{probability} for the behaviour of the boosted and weaker models compared to the fiducial one).

Once a BH is formed, it can grow via three different channels: \textit{hot gas accretion}, \textit{cold gas accretion}, and \textit{mergers} with other BHs. The first growth mechanism is triggered by the continuous accretion of hot gas that surrounds the galaxy \citep{croton2006}. However, since the rate of accretion is orders of magnitude below the Eddington limit, the contribution of this \textit{hot gas accretion} to the BH mass growth is minimal. On the other hand, cold gas accretion is the main channel driving BH growth. This mechanism is triggered by galaxy mergers (where the central BH accumulates a fraction of cold gas from the merged galaxy) and disc instability events (where the BH accretes an amount of cold gas proportional to the mass of stars that have triggered the stellar instability). In both cases, the cold gas available for accretion is assumed to settle in a reservoir around the BH and is progressively consumed according to a two-phase model (as suggested by \citealt{hopkins2009} and first implemented in \texttt{L-Galaxies} by \citealt{bonoli2009, marulli2009}). In the first phase, the central BH consumes a fraction (set to 70\%) of the available gas reservoir following an Eddington-limited growth. In the second phase, the central BH consumes the remaining gas following a quiescent growth regime characterized by progressively smaller accretion rates \citep[for further details on the most recent implementation see][]{izquierdo2020}.

Finally, BHs in the model can also increase their masses through BH coalescences. These do not happen instantaneously, but \texttt{L-Galaxies} includes a detailed model for the dynamical evolution of massive black hole binaries. The evolutionary pathway of these systems can be described by three different stages, as suggested by  \cite{begelman1980}: \textit{pairing}, \textit{hardening} and \textit{gravitational wave} phases. The former takes place just after a galaxy merger when the two BHs are separated at ${\sim}\,\rm kpc$ scales. In particular, the BH hosted by the satellite galaxy experiences dynamical friction that causes its sinking towards the galactic nucleus. In the model, to estimate the time spent by a black hole in the pairing phase, we use the expression by \cite{binney1987} to take into account, among others variables, the mass of the pairing WBH, the velocity dispersion of the satellite galaxy and the position at which the BH is deposited by its satellite galaxy just after the galactic merger (which is considered to be when the satellite galaxy has lost 85$\%$ of its total mass by tidal stripping). Once the dynamical friction phase ends, the satellite BH reaches the galactic nucleus of the primary galaxy and binds with its central BH ($\rm {\sim}\,pc$ scales). As a result, a BH binary is formed. This is when the hardening phase starts and the orbital separation of the BH binary begins to shrink. 
This process is modeled differently according to whether the environment that drives the two BHs to coalescence is a gas-rich or gas-poor environment. Mergers in gas-rich environments require the binary to be surrounded by a gas reservoir with a mass larger than the mass of the binary. 
For mergers in gas-poor environments, it is assumed that the gas reservoir around the BHs is smaller than the total mass of the binary. In these cases, the hardening is caused by the extraction of binary energy and angular momentum through 3-body interactions with background stars that cross the binary orbit. Independently of the environment, the BH binary eventually reaches ${\sim}\,\rm sub\,{-}\,pc$ scales, and the emission of gravitational waves brings the two BHs to the final coalescence in a few Myrs. As described in Section \S \ref{milleniumII}, the baryonic evolution inside \texttt{L-Galaxies} is traced by performing an internal time discretization of 5 -- 20 Myr between the \texttt{Millenium-II} outputs. The evolution of the binary eccentricity and semi-major axis is also traced within that time step by solving their variation via an \textit{adaptative} time step 4th-order Runge-Kutta integrator. Since BH binaries can evolve on  time-scales shorter than the ones provided by \texttt{L-Galaxies} (shorter in the hardening phase and longer in the GW-dominated phase) the variable time step in the integration used in a sub-grid allows us to accommodate the integration of the binary orbit according to its evolution stage. Finally, to trace the binary pathway down to its final coalescence, it is necessary to trace its evolution on scales much smaller than the softening scale of the \texttt{Millennium-II}. To this end, \cite{izquierdo2023} introduced an analytical model in \texttt{L-Galaxies} to trace the sub-pc scales of the BH binaries assuming that the binary hardens according to a stellar distribution (leading the stellar hardening) and the presence of accretion discs (gas hardening) around the BH, together with the emission of GWs.


In some cases, however, the lifetime of a BH binary can be long enough that a third BH which finishes the pairing phase can reach the galaxy nucleus and interact with the binary system. When this happens, the model refers to the tabulated values of \cite{bonetti2018} to determine if the triple interaction leads to the prompt merger of the BH binary or to the ejection of the lightest BH from the system (hereafter tagged as slingshot scenario). In this latter case, the separation of the BH binary that remains in the galactic core is computed following \cite{volonteri2003} and the final eccentricity is select as a random value between [0 -- 1].

\subsubsection{The WBH population} 
\label{formation mechanisms}

Although all the BHs formed in \texttt{L-Galaxies} start their lives inside a galactic nucleus, many different mechanisms can move them away from the heart of the host galaxy.  BHs displaced outside their galaxies nuclei but in bound orbits within the DM halo are tagged as WBHs. \texttt{L-Galaxies} distinguishes and tracks the orbits of three different classes of WBHs according to the physical mechanism that displaced them: \textit{disrupted}, \textit{pairing} and \textit{recoiled}. In the following paragraphs, we outline these types of WBHs and briefly describe their formation mechanism. For a full detailed description of the wandering population we refer the reader to \cite{izquierdo2020} and \cite{izquierdo2022}.

\begin{figure*}
    \begin{center}
    \includegraphics[scale = 0.65]
    {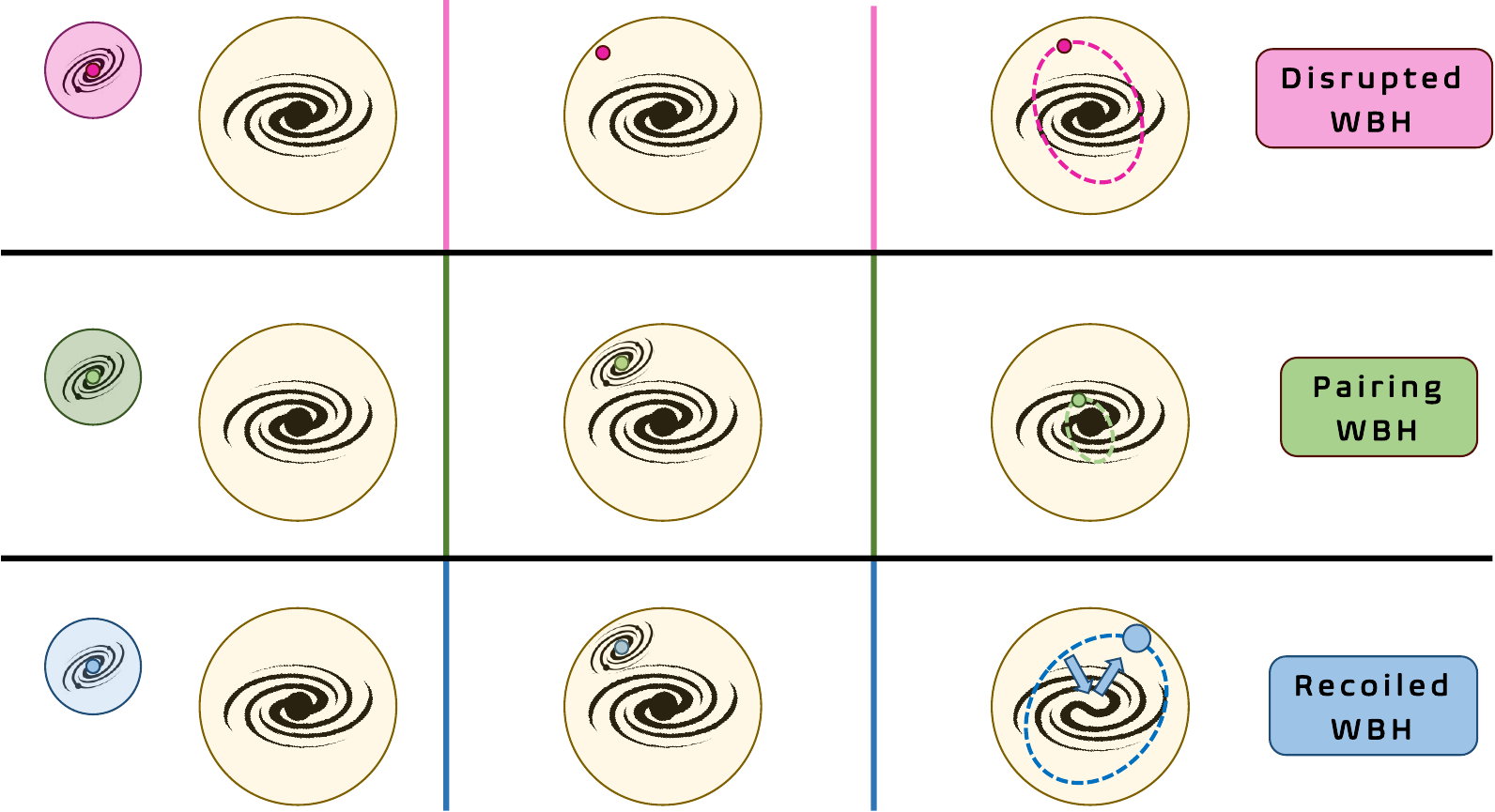} 
     \caption{Schematic representation of the three different WBH formation mechanisms accounted in \texttt{L-Galaxies}. Each row represents a type of WBH, and the columns showcase, from left to right, the steps that lead them to become wanderers. \textbf{First row:} Disrupted WBHs originate from the complete disruption of their host galaxy during their infall towards a central galaxy. Once the BH is deprived of its host galaxy, it is assumed that its accretion disc (in case it was in an accreting phase), is also removed. \textbf{Second row:} Pairing WBHs are infalling BHs that have arrived to the galactic centre of the larger galaxy and are considered by \texttt{L-Galaxies} as being a few mega to giga years to form a binary system with the central BH of the larger galaxy. \textbf{Third row:} Recoiled WBHs are BHs that have also arrived to the galactic centre of the larger galaxy, but then they have continued their sinking process until reaching a distance of $<$100 pc to the central BH. As a result, they have undergone a merging process with its subsequent gravitational recoil. The kick velocity depends on the properties of the two progenitors. If the modulus of the recoil velocity is larger than the escape velocity of the galaxy, the kicked BH is ejected from the host galaxy, starting its wandering phase. The WBH has been depicted bigger in this case to highlight that it is the result of adding the masses of the infalling wanderer and the previous central BH. However, since the central BH mass of high-$z$ galaxies is usually $<$10$^{6}$M$_{\odot}$, recoiled WBHs rarely fit the supermassive BH category. Also, even though a recoil event removes the central BH from the original host galaxy from its galactic nucleus, due to the active merging history of most galaxies and the fact that a pairing WBHs eventually fill this absence, the majority ($>$90$\%$) of recoiled WBHs end up in galaxies that showcase a central BH.}    
    \label{scheme WBH formation}
    \end{center}
    \end{figure*}

\begin{itemize}
    \item \textbf{Disrupted WBHs}: The lifetime of these WBHs starts in the nucleus of small orphan satellite galaxies that at some point in their lives start infalling into a more massive DM halo. As soon as the galaxy becomes an orphan and starts its journey to merge with the central galaxy, \texttt{L-Galaxies} tracks the trajectory of the orphan satellite galaxy at each time step and compares its baryonic (cold gas and stellar gas) density within the half-mass radius to the DM density of the massive halo. If at any point during the infalling, the density from the massive DM halo exceeds that of the satellite, the satellite is completely disrupted, indicating that the satellite is not able to survive the tidal forces exerted by the massive DM halo. As a result, its cold gas (stellar) component is stripped and deposited in the hot gas atmosphere (inter-cluster medium) that surrounds the primary galaxy. Notice that orphan galaxies cannot retain the disrupted (or ejected) WBHs they hosted before losing their DM component. This is because the model assumes that as soon as the galaxy becomes orphan, the disrupted (or ejected) WBHs moving inside the DM halo are also stripped away and incorporated into the halo of the most massive galaxy. Under these circumstances, the newly acquired (disrupted or ejected) WBHs move inside the halo of their new central galaxy and their orbital motion is referred with respect to that galaxy. A schematic view of the formation of these WBHs is presented in the first row of Fig. \ref{scheme WBH formation}. Once a disrupted WBH is formed, \texttt{L-Galaxies} follows its orbit by solving self-consistently its equation of motion. To do so, the code uses the final position and velocity of the galaxy before being disrupted as the starting point of integration of the disrupted WBH population. Then, the code follows their orbit by taking into account the gravity acceleration and the dynamical friction exerted by any of the inter-cluster components (i.e dark matter, stars and hot gas, see Eq.32 of \citealt{izquierdo2020}). \\
    \item \textbf{Pairing WBHs}: 
    Similarly to \textit{disrupted} WBHs, \textit{pairing} WBHs are BHs that originally were hosted in the nuclei of an orphan satellite galaxy. Unlike \textit{disrupted} WBHs, however, the host galaxies of \textit{pairing} WBHs survive the tidal forces exerted over them, meaning that they remain with a higher density compared to the massive DM halo throughout the whole infall. During this process, the satellite galaxy manages to reach the primary galaxy and merge with it. Notice that, unlike the disrupted and ejected WBHs, in case the satellite galaxy had pairing WBHs prior to becoming orphan (coming from mergers with other galaxies when it was still a central galaxy), these are still able to be retained by the galaxy during its lifetime as orphan. This is because these WBHs are assumed to be moving inside the galaxy, thus can only be unbound by a full galaxy disruption. Once the orphan galaxy survives the tidal stripping all the way towards the primary galaxy and the merger is completed, the pairing WBHs hosted in the orphan galaxy are deposited inside the central one. As a result, the BH population of the orphan satellite galaxy is deposited in the outskirts of the newly formed galaxy and progressively sink towards the nucleus of the merger remnant thanks to the dynamical friction force exerted by the stellar component. A schematic view of the formation of these WBHs is presented in the second row of Fig. \ref{scheme WBH formation}. Unlike \textit{disrupted} WBHs, we do not follow the orbit of \textit{pairing} WBHs. Instead, we assume that they will reach the galactic centre after spending a time $t_{\rm dyn}$ in the paring phase. This time is computed according to the dynamical friction equation of \cite{binney1987} which assumes, for simplicity, a spherical isothermal galaxy. Regarding the evolution of the \textit{pairing} WBHs, we assume a simple model in which the BHs spiral towards the galactic nucleus inside an isothermal host on a near-circular orbit. This assumption predicts that the radius of the orbit should decay linearly in time. Consequently, the radial position of the BH inside the galaxy is updated at each time step of \texttt{L-Galaxies} by multiplying the previous distance by the factor $(1\,{-}\,\Delta t/t_{\rm dyn})^{1/2}$, where $\Delta t$ is the time since the BH started the pairing phase \citep[see similar approach used in][to follow the orbit of satellite galaxies inside DM haloes]{guo2011}. Finally, in this work, the minimum allowed distance with respect to the galactic centre to consider a pairing WBH as a wandering object is $100\,\rm pc$. Pairing WBHs at distances below this threshold are assumed to no longer wander since they will form in the next time step of \texttt{L-Galaxies} (a few Myr) a binary system with the central BH of the central galaxy. \\
    
    \item \textbf{Recoiled WBHs}: This type of WBHs follows the same formation pathway of pairing WBHs; a satellite galaxy merges with a central galaxy after surviving the tidal forces applied by a DM halo and the BH of the satellite becomes a pairing WBH. After a certain time (from mega to giga years), the pairing WBH reaches the nucleus of the galaxy and forms a BH binary that evolves through the hardening and gravitational wave phase. When these two phases are over, the two BHs in the binary system coalesce. Due to the merger and conservation of linear momentum, the remnant BH receives a kick (determined according to \citealt{Lousto2012}). If the magnitude of the velocity characterizing such a kick is large enough, the BH escapes from the centre of the galaxy and is deposited inside the DM halo that hosts the galaxy. A schematic view of the formation of these WBHs is presented in the last row of Fig. \ref{scheme WBH formation}. Once the BH is ejected from the galaxy, \texttt{L-Galaxies} follows its orbit in the same way as described for the \textit{disrupted} WBHs. Notice that the slingshot scenario presented in the \S \ref{formation and growth} forms wandering BHs similar to the recoiled ones. In these cases, the BH kick that ejects the BH is not caused by the merger of two BHs but by scattering after three-body interaction (the two BHs in the binary and the intruder pairing WBH). For these cases, the orbits are followed in the same way as presented in the recoiled case and, for convenience, they are tagged as recoiled as well, since they only represent $\ll$1$\%$ of the total WBH population.
\end{itemize}

As per the wanderers themselves, disrupted and recoiled WBHs are treated as not having an accretion disc. In the case of disrupted WBHs, their accretion disc is removed, and in the case of recoiled WBHs, their accretion disc is also neglected since it is unclear how massive the retained disc would be after a recoil (see \citealt{blecha2011}). In the case of pairing WBHs, they are able to increase their gas reservoir during the galaxy-galaxy interaction and therefore experience accretion enhancements. However, no feedback effects are considered during this phase. In summary, once dislodged from their satellite galaxy, WBHs of any kind have no direct impact on the evolution of the new host galaxy through feedback mechanisms. The only time that a direct impact takes place (in the form of radio feedback) is when a pairing WBH reaches the galactic centre of the primary galaxy and becomes its new central BH, but in this scenario, the BH is no longer considered a wanderer. Finally, the model does not consider any dynamical interaction between WBHs and infalling galaxies; they are just dominated by the DM halo potential.

We also stress that as the redshift gets lower and the merger history of galaxies gets more complex, satellite galaxies deposit not only their central BHs in the DM haloes of larger galaxies but also the WBH population that is orbiting them as a consequence of the three main wandering events we just mentioned. Consequently, it is expected that the majority of the WBHs orbiting low redshift galaxies are not generated in situ but inherited from other galaxies (see e.g Fig. 8 of \citealt{izquierdo2020}). For instance, we have checked that $\sim$70$\%$ of the WBHs detected in a MW-type galaxy at $z$ = 0 are actually inherited from the wandering populations of previous galaxies.

\subsection{Milky Way-type galaxies}
\label{milky way}
Similarly to previous works \citep[see, e.g,][]{licquia2013,murphy2022}, we have defined MW-type galaxies as those fulfilling the following constraints:

\begin{itemize}
    \item Have a stellar mass within [2$\times$10$^{10}$ -- 8$\times$10$^{10}$] M$_{\odot}$ 
    \vspace{1mm}
    \item Have a DM halo with mass [6$\times$10$^{11}$ -- 3$\times$10$^{12}$] M$_{\odot}$ 
    \vspace{1mm} 
    \item Have a star formation rate in the range [0.5 -- 2.5] M$_{\odot}$yr$^{-1}$
    \vspace{1mm}
    \item Have a bulge to total stellar mass ratio within [0.1 -- 0.5]
    \vspace{1mm}
    \item Are central galaxies of their subhaloes    
\end{itemize}

\subsection{Leftover seeds}
\label{leftover seeds}
Although plenty of literature has discussed the characterization of leftover seeds (e.g, \citealt{rashkov2013, ferrara2014, greene2019, mezcua2019, mezcua2020feedback}), no definitive consensus has been reached regarding their properties. Typically, they are theorized as BHs that have had little to no growth in their mass since their time of formation. In this work we will tag as a leftover seed any BH that has not outgrown more than twice its original mass. Since the range of masses for newly seeded BHs is within [10$^{2}$ -- 10$^{4}$] M$_{\odot}$, the mass range for leftover seeds at $z$ = 0 will be [10$^{2}$ -- 2 $\times$ 10$^{4}$] M$_{\odot}$.

\section{Results}
\label{results}
In this section, we present the results obtained from utilizing \texttt{L-Galaxies} to characterize the WBH population in MW-type galaxies at the current redshift (see \S \ref{WBH population}). Additionally, we delve into the merger history of the WBH population, establishing a connection between their current properties and the redshift at which they became wanderers (see \S \ref{WBH history}). In \S \ref{subsection leftover seeds} we focus on the leftover seed population of WBHs, characterizing it in the same way as the previous subsections and comparing it with WBHs that instead managed to substantially grow in mass. Finally, we conduct a comprehensive review of the \texttt{L-Galaxies} seeding prescription and its effects on the stellar mass of the galaxies that hosted the WBHs before they became wanderers (see \S \ref{L-galaxies seeding prescription}).

\subsection{The population of WBHs in MW-type galaxies}
\label{WBH population}
\begin{figure}
\begin{center}

\subfloat{%
  \includegraphics[scale = 0.6]{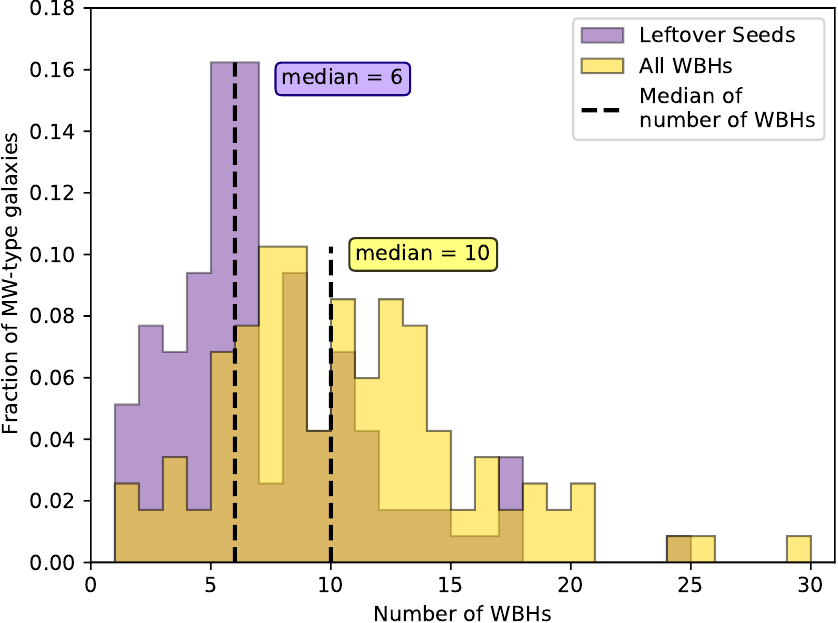}%
}

\subfloat{%
  \includegraphics[scale = 0.6]{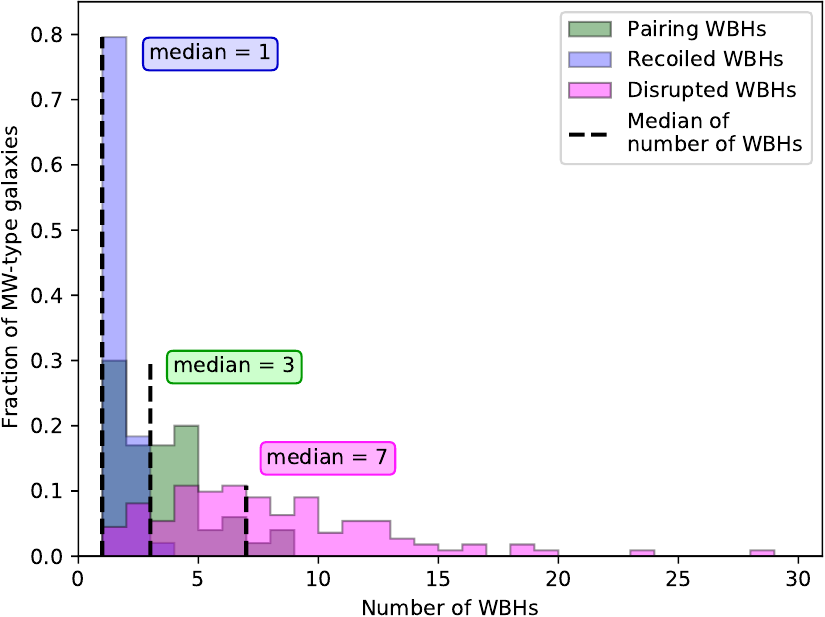}%
}

\caption{Distribution of MW-type galaxies as a function of the number of WBHs that they are hosting. The dashed lines indicate the expected number of WBHs of a given population. \textbf{Top.} Fraction of MW galaxies hosting any kind of WBH (yellow) compared to the fraction of MW galaxies hosting leftover seeds (violet). \textbf{Bottom.} Fraction of MW galaxies hosting WBHs that have undergone a process of disruption (magenta), an ejection (blue), or that are in the process of forming a binary system with their central BH (green). See more details regarding the formation mechanisms of the WBHs in \S \ref{formation mechanisms}.}
\label{BH occupation fraction}

\end{center}
\end{figure}

The \texttt{L-Galaxies} fiducial model run on the \texttt{MR-II} offers information on $\sim$27.000 galaxies at $z$ = 0. After applying our constraints for MW-type galaxies (see \S \ref{milky way}), we end up with a sample of 117 galaxies. All of these MW-type galaxies host at least one WBH, with the predicted number of WBHs per galaxy being 10$_{-3}^{+5}$\footnote{All predicted numbers are computed as the median of the distributions, and the errors as the 20th and 80th-percentiles, respectively.} (see the yellow histogram in the top panel of Fig. \ref{BH occupation fraction}).

We also study the statistics of WBHs depending on their formation channel (see \S \ref{formation mechanisms}). The results are presented in the bottom panel of Fig \ref{BH occupation fraction}. Of the 117 MW-type galaxies, $\sim$97$\%$ of them host at least one disrupted WBH, with the expected number being 7$_{-3}^{+5}$ for this type of wanderer; $\sim$85$\%$ of MW-type galaxies host one or more pairing WBH, with the expected number being 3$_{-2}^{+1}$; and $\sim$42$\%$ of MW-type galaxies host one or more recoiled WBH, with the expected number being 1$_{-0}^{+1}$. The higher expected number of disrupted WBHs compared to the other two families might be explained by the fact that pairing and recoiled WBHs result from their host satellite galaxies sinking into a massive DM halo and merging with a larger galaxy. Since this entire process takes more time than a disruption event, which can occur as soon as a satellite galaxy loses most of its mass, disrupted WBHs end up being the most abundant population.

\begin{figure}
    \begin{center}
    \includegraphics[scale = 0.6]{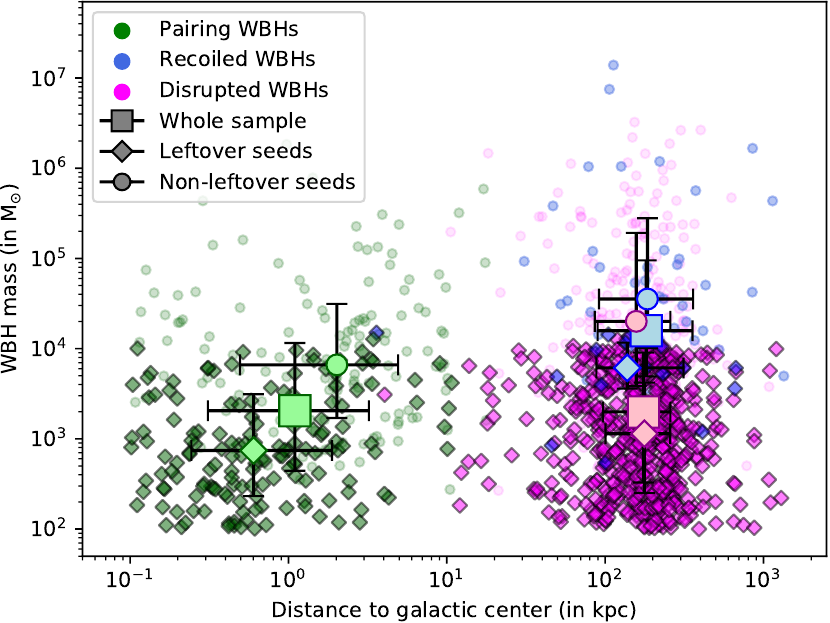} 
     \caption{Mass distribution of the WBHs in MW-type galaxies as a function of their distance to the galactic centre. The WBHs are color-coded according to whether they are in the process of forming a binary system with the central BH of their galaxy (green), whether they have undergone a recoil event (blue) or if they have undergone a disruption event (magenta). The diamond-shaped points represent the population of leftover seeds while the dots their non-leftover counterpart. The big symbols on top of the points indicate the median values of WBH mass and distance to galactic centre for the total sample of WBHs (square), the leftover seed subsample (diamond) and the non-leftover seed subsample (circle). The errors represent the 20th and 80th percentile of each distribution.}
      
    \label{BH mass distance}
    \end{center}
    \end{figure}

Fig. \ref{BH mass distance} presents a view of the mass distribution of the WBH population (points and diamonds) as a function of its distance to the galactic centre, color-coded according to each of the three WBH types. We find that pairing WBHs are expected to be found at 1.1$_{-0.8}^{+2.1}$ kpc from the galactic centre and to have a mass of $\bigl($ 2.0$_{-1.6}^{+9.5}$ $\times$ 10$^{3}$ $\bigr)$ M$_{\odot}$ (Fig. \ref{BH mass distance}, green square); disrupted WBHs are expected to be found at 174$_{-78}^{+84}$ kpc from the galactic centre and to have a mass of $\bigl($ 2.0$_{-1.7}^{+7.0}$ $\times$ 10$^{3}$~$\bigr)$ M$_{\odot}$ (Fig. \ref{BH mass distance}, red square); and recoiled WBHs are expected to be found at 182$_{-92}^{+174}$ kpc from the galactic centre and to have a mass of $\bigl($ 1.6$_{-11}^{+79}$ $\times$ 10$^{4}$ $\bigr)$ M$_{\odot}$ (Fig. \ref{BH mass distance}, blue square). This picture is coherent with the evolutionary stage of our population of WBHs. Pairing WBHs are in the process of sinking towards their galactic nucleus after the galaxy interaction whereas disrupted and recoiled WBHs are embedded in the DM haloes in orbits that can linger in the halo outskirts thanks to large gravitational wave recoils and galaxy disruption processes occurring primarily at the halo outskirts. Delving into the pairing population, these type of WBHs are most likely found in the galactic disc, where the stellar density is higher and the dynamical friction allows for a more efficient sinking. This, in turn, exacerbates the detection of potential compact stellar clusters bound to them, as stellar overdensities in the disc are easily confused with open clusters.

On the other hand, we see that recoiled WBHs showcase higher mass values than the other WBH families. This is expected since recoil events are the result of a coalescence that includes a central BH, and central BHs have generally had some time to grow before the merger. A small fraction (<0.1$\%$) of the WBH population in MW-type galaxies does not experience disruption, pairing or recoiled events. This is the case for the \textit{slingshot} wanderers. Slingshot wanderers are the result of a triple BH interaction at the centre of a galaxy, which leads to the ejection of the lightest BH (see \S \ref{formation mechanisms}). Of the 1310 WBHs scattered among 117 MW-type galaxies, only one WBH was tagged as having undergone a slingshot event. When compared to the overall sample of galaxies, slingshot events are slightly more probable, occurring in $\sim$0.3$\%$ of all WBHs. 

Combining these results with previous observational works (e.g., \citealt{oleary2012, greene2021}), we can derive a more holistic interpretation to the nature of WBHs. Since one potential method to detect WBHs in our galaxy could reside in the compact cluster of stars that are still bound to them, the search for hypercompact (15$\arcsec$ -- 30$\arcsec$) stellar overdensities could be one of the most promising paths to reveal this population \citep{oleary2009}. However, most used surveys like SDSS \citep{oleary2012} and Gaia $\&$ DECaLS \citep{greene2021} have a limiting band depth that only allows for the study of well-defined stars in color magnitude diagrams up to $\sim$25 kpc. This limits the search for WBHs to almost the pairing population exclusively, and given the high-density environment of the galactic disc, hypercompact clusters can easily be mistaken for regular open clusters. Additionally, we must take into account the spatial limitation of these surveys; the DECaLS database covers about 1/5 of the sky and SDSS about 1/3. This further reduces the possibility of finding a member of the pairing WBH population, which in our model represents $\sim$25$\%$ of the overall predicted population of wanderers ($\sim$10). We will use these results to also interpret the upcoming observational analysis of Untzaga et al. (in prep).

Other observational techniques to detect these pairing wanderers include microlensing events when they pass along the line of sight of a background star \citep{toki2021}, radio to X-ray detection thanks to telescopes like AXIS that have improved their detectability in high-density environments \citep{seepaul2022, pacucci2024} and indirect dynamical evidence when the wanderer is electromagnetically invisible \citep{kiziltan2017}.

\begin{figure}
    \begin{center}
    \includegraphics[scale = 0.6]{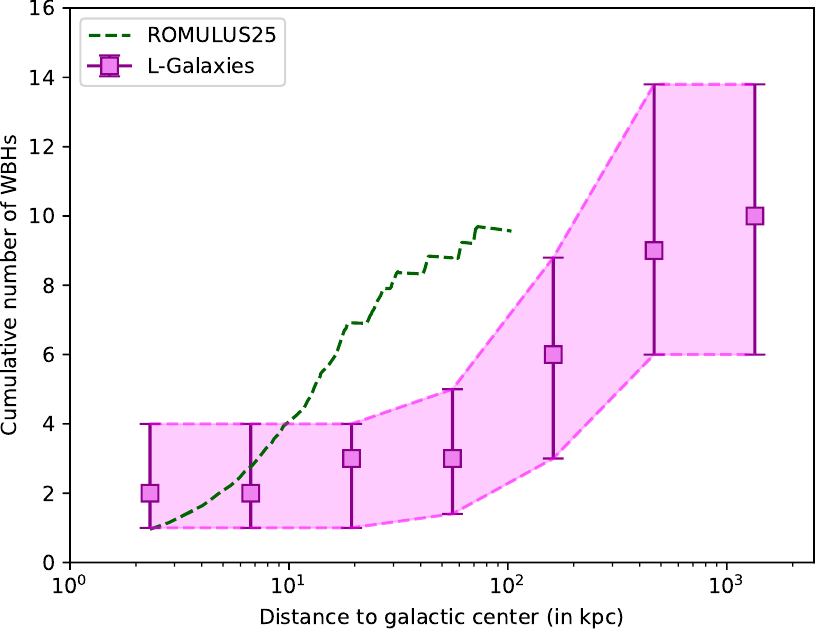} 
     \caption{Expected number of WBHs that MW-type galaxies host within a certain distance to their galactic centre, with the magenta shaded region indicating the 80th and 20th percentiles. The green dashed line represents the median cumulative number of wandering SMBHs as a function of distance from halo centre for all MW-Mass haloes in the \texttt{ROMULUS25} hydrodynamic simulation by {\protect \cite{tremmel2018}}.}
        
    \label{comparison Romulus}
    \end{center}
    \end{figure}

In Fig. \ref{comparison Romulus} we compare our results with the \texttt{ROMULUS25} hydrodynamic simulation from \cite{tremmel2018}. We find a good agreement in the total number of WBHs expected in a MW-type galaxy ($\sim$10), and especially in the number of WBHs within 5 kpc of the galactic centre ($\sim$2), which in our case corresponds to the pairing WBH population. However, we find the radial distribution of the non-pairing WBH population to be far more extended than predicted in \cite{tremmel2018}, with $\sim$60$\%$ of all WBHs being farther than 100 kpc from their galactic centre (see Fig. \ref{comparison Romulus}). This is probably due to the fact that the initial seeding mass for BHs in \texttt{ROMULUS25} is 10$^{6}$ M$_{\odot}$, while \texttt{L-Galaxies} uses a lighter distribution (100 -- 10$^{4}$) M$_{\odot}$. Since the WBH population in \texttt{ROMULUS25} has larger masses, the sinking process takes place more efficiently, and therefore WBHs end up  closer to the galactic centre. It is worth mentioning that \texttt{ROMULUS25} does not consider multi-body BH interactions nor gravitational wave recoil from BH mergers as wanderer-triggering events for their BHs. In \texttt{ROMULUS25}, a BH is tagged as a wanderer if it is further than 700 pc from the centre of its host halo, but no recoiled WBH population is contemplated.

\begin{figure}
    \begin{center}
    \includegraphics[scale = 0.6]{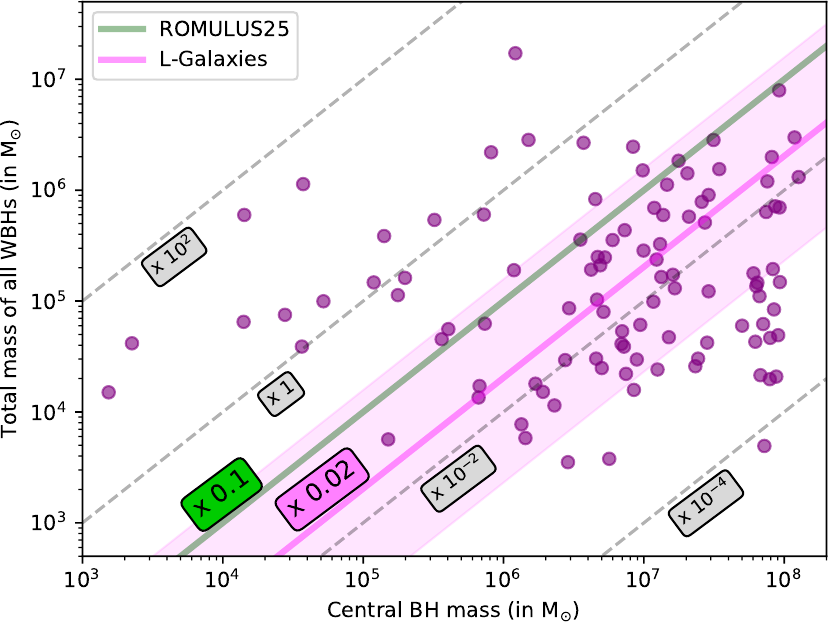} 
     \caption{The magenta dots show the relation between the total WBH mass of the MW galaxies as a function of the mass of their central BHs. Only those MW-type galaxies hosting a central BH are shown (103 out of 117). The magenta line indicates the median of the ratio for MW-type galaxies in our simulation, while the shaded area indicates the 20th and 80th percentiles. We find that WBHs in a MW-type galaxy have a combined total mass that represents a $\sim$2$\%$ with respect to the mass of their central BH. The green line indicates the same ratio in the hydrodynamic simulation \texttt{ROMULUS25}, as extracted from {\protect \cite{ricarte2021}}, where they find a ratio of $\sim$10$\%$. Grey dotted lines have been added for visual reference, showing ratios of mass separated by two orders of magnitude.}
        \label{central ratio}     
    \label{ratio black holes}
    \end{center}
    \end{figure}

In Fig. \ref{central ratio} we compute the ratio between the total mass of all WBHs in each MW-type galaxy with respect to the mass of their respective central BH, and compare our results with the analysis of the \texttt{ROMULUS25} simulation from \citealt{ricarte2021}. We find that in \texttt{L-Galaxies} WBHs account for 2$_{-1.8}^{+14}$$\%$ of the local non-stellar BH mass budget of a MW-type galaxy, which is lower than the $\sim$10$\%$ found by \texttt{ROMULUS25}. The disagreement between these two values is most likely due to the minimum WBH mass in \texttt{ROMULUS25} being 10$^{6}$ M$_{\odot}$. In comparison, the minimum WBH mass accounted for in \texttt{L-Galaxies} is $\sim$100 M$_{\odot}$, which makes for a more \textit{centralized} distribution of the BH mass budget in the galaxy. 

Note that some of the galaxies in our sample have a central BH with a mass significantly lower than the ones predicted by the scaling relations derived from massive galaxies. This is in part because the inclusion of recoil events causes central BHs to be off-centre (thus not growing) for even extended periods of time, and partially because in some cases, the galaxy incorporated as central BH is one of the infalling satellites (see \citealt{izquierdo2020}).

\subsection{The history of WBHs in MW-type galaxies}
\label{WBH history}
\begin{figure*}
     \centering
     \begin{subfigure}[b]{0.33\textwidth}
         \centering
         \includegraphics[width=\textwidth]{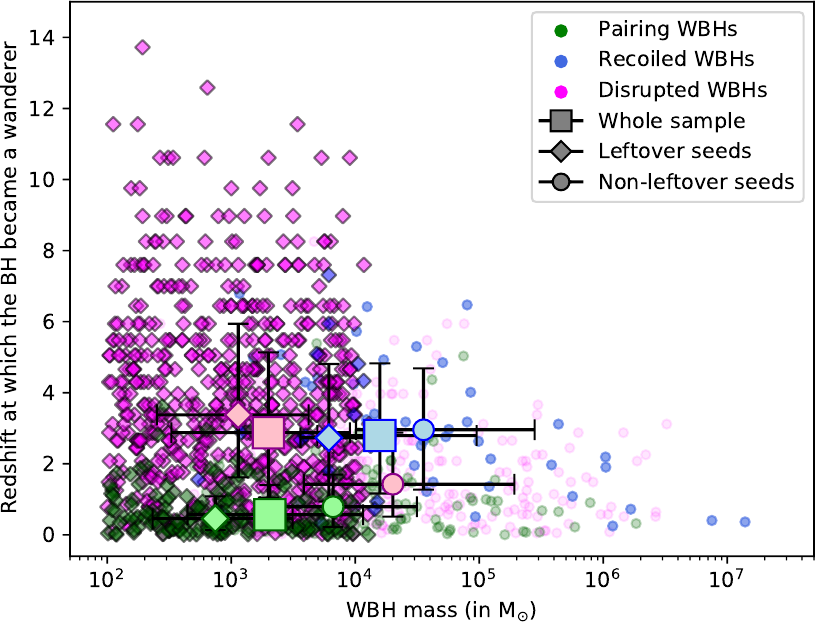}
     \end{subfigure}
     \hfill
     \begin{subfigure}[b]{0.33\textwidth}
         \centering
         \includegraphics[width=\textwidth]{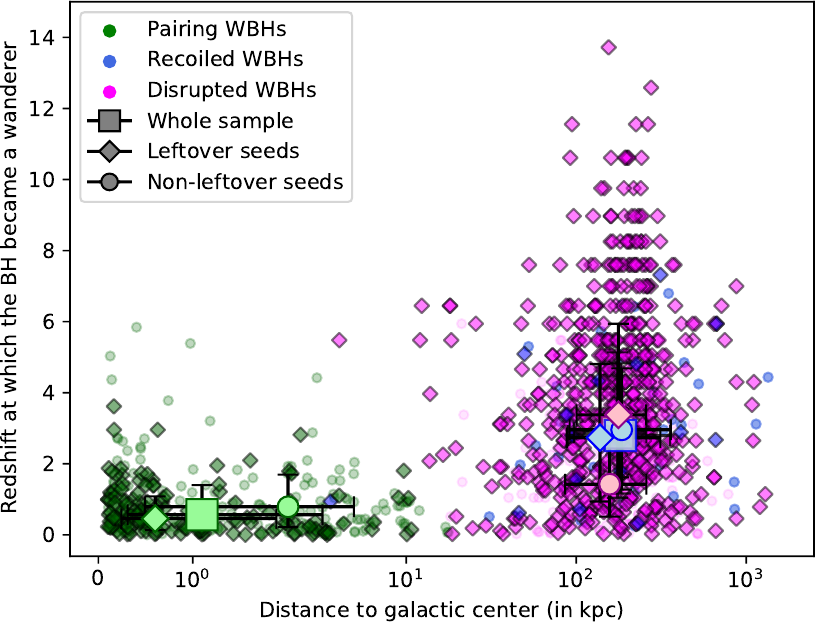}
     \end{subfigure}
     \hfill
     \begin{subfigure}[b]{0.33\textwidth}
         \centering
         \includegraphics[width=\textwidth]{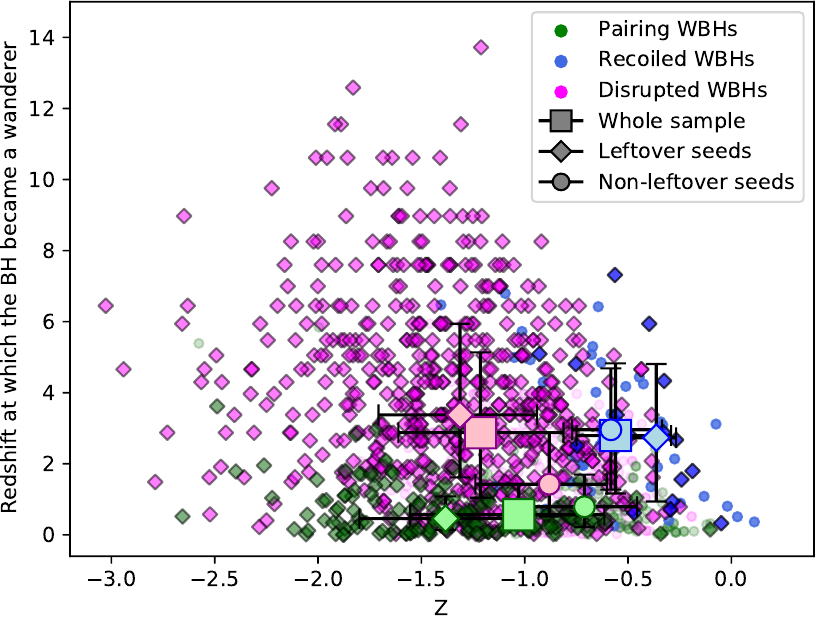}
     \end{subfigure}
        \caption{Redshift at which WBHs first became wanderers versus their current mass (left), distance to galactic centre (middle), and the metallicity of the galaxy where the wandering event takes place (right). The colors and symbols used are the same as in Fig. \ref{BH mass distance}. Note that in the middle Figure, the blue and red squares representing the whole sample medians of the recoiled and disrupted populations respectively, overlap in the same space.}
        \label{WBH history triple plot}. 
\end{figure*}

The study of the merger history of the WBH population is crucial to improve our understanding of these objects in the current Universe. Fig. \ref{WBH history triple plot} shows some of the properties that BHs (points and diamonds) showcase at the time they become wanderers. We find the earliest event of disruption to occur at $z$ $\sim$ 14, while in the case of recoil events this happens at $z$ $\sim$ 7, and in the case of pairing at $z$ $\sim$ 6. Until $z$ $\sim$ 5, the formation channel for WBHs is heavily dominated by disruption events. These early disruption events result in wanderers that do not continue to grow beyond the upper mass limit of the initial BH seeding given by \texttt{L-Galaxies}, hence being all below 10$^{4}$ M$_{\odot}$. 
When focusing exclusively on disruption events, early ones ($z$ $\geq$ 5) take place in galaxies with lower metallicities (-1.47 compared to -1.12 for disruption events at $z$ $<$ 5) and are also linked to wanderers that end up wandering in farther orbits at current redshift (193 kpc compared to 169 kpc). This is relevant when trying to detect WBHs in the MW through observational means. As discussed by \cite{greene2021}, $\sim$100 BHs with masses 10$^{3}$ -- 10$^{5}$ M$_{\odot}$ are expected to be wandering in the MW surrounded by $\sim$1 pc size stellar clusters containing 500~--~5000 stars. Even though in this work we find a smaller number of putative WBHs, the values found for galactic metallicity in the galaxies of origin can be used to tighten the search for stellar clusters that might still be bound to the wanderers, and that would have once belonged to that same galaxy of origin. Constraining the metallicity values is also useful when proceeding with the isochrone fitting of the stellar population's CMDs (e.g., to verify the old nature of these kind of clusters).

Observationally, pairing WBHs could be differentiated from the rest of wanderers by their unique radial distribution close to the galactic centre (see Fig. \ref{BH mass distance}), and also their dynamical interactions with the stellar medium. Disrupted and recoiled WBHs could be differentiated between each other by the metallicity of the stellar populations bound to them. In this case, recoiled WBHs should be surrounded by stellar populations with higher metallicity due to potential star capture during the infall process towards the galactic core of a central galaxy).

\subsection{The population of leftover seeds in MW-type galaxies}
\label{subsection leftover seeds}
We study the subsample of WBHs that have not outgrown more than two times their original seeding mass (see \ref{leftover seeds}). We find that $\sim$67$\%$ of the WBH population in a MW-type galaxy at $z$ = 0 satisfies this constraint. We also find that all of the MW-type galaxies host at least one or more leftover seed, with the median number per galaxy being 6$_{-2}^{+4}$ (see Fig. \ref{BH occupation fraction}, Top).

Of the 117 MW-type galaxies, $\sim$93$\%$ of them host at least one disrupted leftover seed, with the expected number being 5$_{-2}^{+3}$ for this type of seed; $\sim$72$\%$ of MW-type galaxies host one or more pairing leftover seeds, with the expected number being 1$_{-0}^{+2}$; and $\sim$12$\%$ of MW-type galaxies host one or more recoiled leftover seeds, with the expected number being 1$_{-0}^{+0}$. When compared with the overall sample of WBHs, it seems that leftover seeds are less affected by recoiled events. This can be explained by the fact that recoiled WBHs are generally the most massive of the three families, and since leftover seeds are defined as having a limit in growth, it is expected that fewer of them will be tagged as recoiled leftover seeds.

The diamond-shaped points in Fig. \ref{BH mass distance} showcase the mass distribution of the leftover seed wandering population as a function of its distance to the galactic centre, color-coded according to each of the three formation mechanisms. We find that pairing seeds are expected to be found at 0.6$_{-0.4}^{+1.3}$ kpc from the galactic centre and to have a mass of $\bigl($ 0.7$_{-0.5}^{+2.4}$ $\times$ 10$^{3}$ $\bigr)$ M$_{\odot}$; disrupted seeds are expected to be found at 177$_{-76}^{+81}$ kpc from the galactic centre and to have a mass of $\bigl($~1.1$_{-0.9}^{+3.1}$ $\times$ 10$^{3}$ $\bigr)$ M$_{\odot}$; and recoiled seeds are expected to be found at 138$_{-50}^{+175}$ kpc from the galactic centre and to have a mass of $\bigl($ 6.1$_{-2.5}^{+5.5}$ $\times$ 10$^{3}$ $\bigr)$ M$_{\odot}$. When compared to the sample of non-leftover seeds, disrupted seeds are $\sim$13$\%$ farther from the galactic centre; recoiled seeds are $\sim$25$\%$ closer to the galactic centre; and pairing seeds are $\sim$70$\%$ closer to the galactic centre. 

As per the merger history, we find that the majority of WBHs in MW-type galaxies (both leftover seeds and non-leftover seeds) that became wanderers at high redshift are most likely tagged as disrupted (see Fig. \ref{WBH history triple plot}). This is probably due to pairing and recoiled WBHs being the result of a galaxy merger that requires a satellite galaxy to sink into a massive DM halo, therefore occurring at a later time in the simulation. 

We also compare the expected wandering event redshift of a leftover seed to its non-leftover counterpart. We find that disrupted seeds become wanderers at a higher redshift ($\sim$3.4 vs $\sim$1.4); pairing seeds become wanderers at a lower redshift ($\sim$0.4 vs $\sim$0.8); and recoiled seeds and non-seeds have very similar redshifts of formation. In the case of the disrupted population, this is most likely due to BHs drastically slowing their growth rate once becoming wanderers. That would explain why the majority of leftover seeds experience their disruption event at a higher redshift compared to their non-leftover counterparts, which have had more time to continue their growth. In the case of the pairing population, leftover seeds are tagged as \textit{pairing} at a lower redshift probably due to being hosted by galaxies with a lower mass compared to their non-leftover counterparts. This means that pairing seeds would sink slower into the galactic core due to experiencing less dynamical friction. Finally, in the case of the recoiled population, especially the recoiled seeds, we find very low numbers, as expected. 

Finally, We compare the leftover and non-leftover populations in terms of the metallicity of the galaxy that originally hosted them (Fig. \ref{WBH history triple plot}, right). We find that disrupted and pairing leftover seeds come from more metal-poor galaxies than non-leftover seeds (--1.3 vs --0.9 in the case of disrupted ones and --1.4 vs --0.7 in the case of pairing ones). This is most likely due to old, low-mass, low-metallicity galaxies being the ones to host BHs that have little to no growth during their lives. In the case of the recoiled WBHs (both leftover seeds and non-leftover seeds), the metallicity Z does not correspond to the galaxy that originally hosted the BH, but rather the one where the coalescence process took place. We find that, in contrast to disrupted and pairing leftover seeds, recoiled seeds come from more metal-rich environments (--0.4 vs --0.6). Despite $\sim$70$\%$ of the recoiled seed population being related to galactic environments with metallicity Z~>~--0.5, only $\sim$30$\%$ of their non-leftover counterparts are found in that regime. From an observational standpoint, metallicity values associated with leftover seeds in low-density environments (non-pairing seeds), are key in our effort to identify potential stellar populations bound to them, as these populations would likely exhibit metallicity values similar to those of their host galaxies. By narrowing the range of metallicities, we can generate more accurate mock isochrones to aid in the CMD fitting process and potentially distinguish leftover seed candidates from regular WBHs.

\subsection{The impact of BH seeding efficiency on the population of WBHs}
\label{L-galaxies seeding prescription}
We study the effects of varying the initial seeding prescription of \texttt{L-Galaxies} on the merger history of the WBH population. We do so by running the model on two new seeding prescriptions: a boosted one where the seeding probability $\mathcal{P}$ is doubled with respect to the fiducial model, and a weaker one where the probability is halved (see \S \ref{BH seeding} for more details). 

\subsubsection{The global population of WBHs in MW-type galaxies}

\begin{figure*}
     \centering
     \begin{subfigure}[b]{0.33\textwidth}
         \centering
         \includegraphics[width=\textwidth]{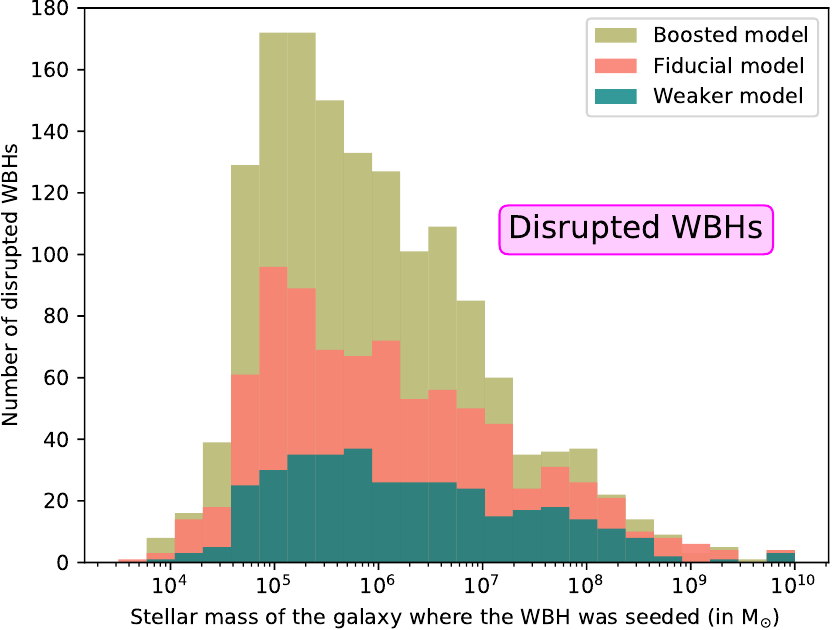}
     \end{subfigure}
     \hfill
     \begin{subfigure}[b]{0.33\textwidth}
         \centering
         \includegraphics[width=\textwidth]{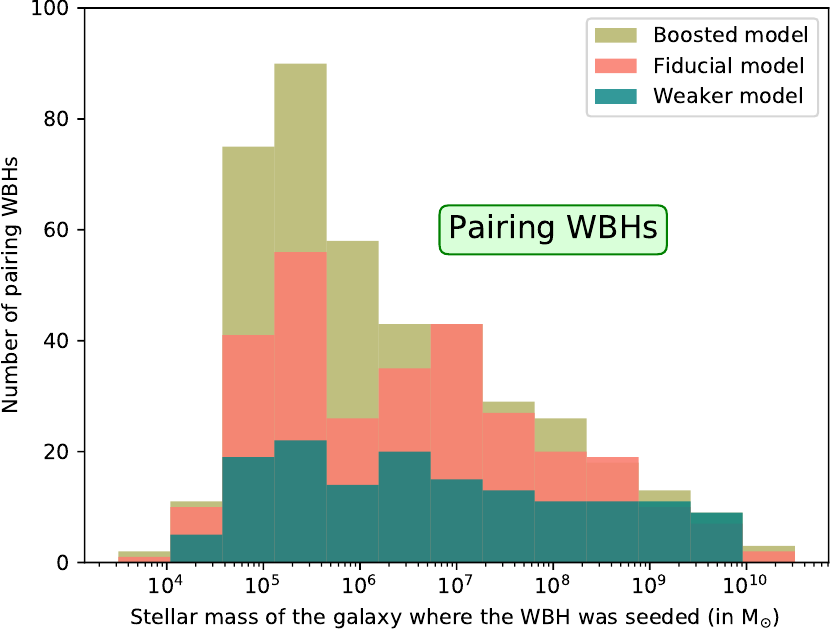}
     \end{subfigure}
     \hfill
     \begin{subfigure}[b]{0.33\textwidth}
         \centering
         \includegraphics[width=\textwidth]{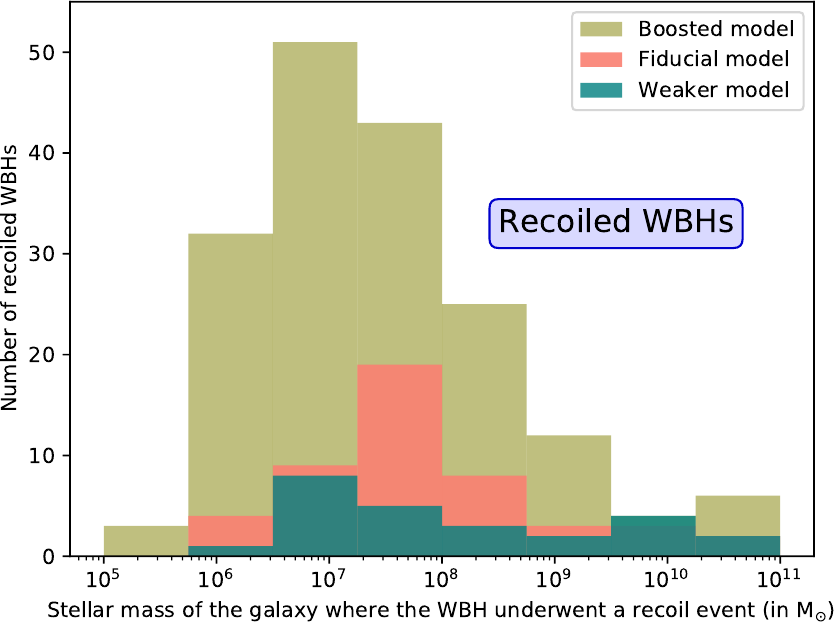}
     \end{subfigure}
        \caption{Number of disrupted (left), pairing (middle), and recoiled (right) WBHs as a function of the stellar mass of the galaxy that hosted them originally. In the case of the recoiled WBHs (right), the x-axis does not represent the stellar mass of the galaxy where the WBHs were first seeded, but rather the galaxy with which the BH-BH fusion and the subsequent recoil took place. The fiducial model is shown in red, while a version with boosted BH seeding is shown in olive and a weaker one in blue (see \S \ref{BH seeding} for more details).}
        \label{WBH seeding prescription triple plot}. 
\end{figure*}

We do not show the plots here to not overcrowd the paper, but we compared the WBH population in the boosted and weaker models to the fiducial one. In the boosted model, a typical MW galaxy hosts 50$\%$ more WBHs, with masses $\sim$10$\%$ smaller for the non-pairing WBHs and $\sim$25$\%$ smaller for the pairing ones. In the weaker model, a typical MW galaxy hosts 50$\%$ fewer WBHs, with masses $\sim$20$\%$ greater for the non-pairing WBHs and $\sim$85$\%$ greater for the pairing ones. In terms of the percentage of WBHs of each class, while the fiducial model shows a ratio of $\sim$70/25/5, corresponding to disrupted, pairing and recoiled WBHs respectively, the boosted model shows a ratio of $\sim$71/20/9 and the weaker model a ratio of $\sim$68/28/4. This indicates that, while the presence of disrupted and recoiled WBHs increases in models with higher seeding probabilities, the presence of pairing WBHs decreases. One possible explanation may lie in the fact that models with boosted BH seeding tend to favor earlier galactic mergers, giving BHs from satellite galaxies more time to reach the centre of larger galaxies, where they can be absorbed or ejected by their central BHs. Regarding the locations of WBHs within the galaxy, we find that the distance to the galactic centre of all the wandering populations is almost invariant to the seeding probability of the model, with differences <10$\%$ (with the exception of the recoiled population in the boosted model, located $\sim$20$\%$ farther than the fiducial model, most likely due to interactions with more massive central BHs).

We also study the seeding probability's effect in the WBH population's merger history. Fig. \ref{WBH seeding prescription triple plot} showcases the distribution of the number of WBHs per bin of stellar mass of the galaxy that hosted them when becoming wanderers. As expected, a higher seeding probability in the early Universe results in a greater amount of galaxies hosting the type of BHs that will later become wanderers in MW-type galaxies. However, we see an inverted correlation between the expected stellar mass of the host galaxy and the seeding probability of its model. This correlation can be observed in the disrupted WBH population, which is the largest of the three samples, where host galaxies in the boosted model have an expected stellar mass of (5.8 $\times$ 10$^{5}$) M$_{\odot}$, while the fiducial and weaker models showcase bigger ones (8.0 $\times$ 10$^{5}$ M$_{\odot}$ and 1.2 $\times$ 10$^{6}$ M$_{\odot}$, respectively). The pairing WBH population mirrors this behavior with the boosted, fiducial and weaker models having an expected stellar mass for the host galaxies of 8.9 $\times$ 10$^{5}$ M$_{\odot}$, 2.6~$\times$~10$^{6}$~M$_{\odot}$ and 3.8 $\times$ 10$^{6}$ M$_{\odot}$ respectively. The same applies for the recoiled WBH population, with the boosted, fiducial and weaker models being, in this case, 1.8 $\times$ 10$^{7}$ M$_{\odot}$, 5.5~$\times$~10$^{7}$~M$_{\odot}$ and 6.2~$\times$~10$^{7}$~M$_{\odot}$ respectively. However, the overall higher expected mass of the three models in the recoiled population (at least when compared to the other two populations) is due to \texttt{L-Galaxies} not considering the stellar mass of the galaxy where the WBH was originally seeded (as it was the case for the disrupted and pairing populations), but rather the central galaxy where the coalescence and subsequent recoil took place.

To explain the inverse correlation between the seeding probability and the expected stellar mass of galaxies during the merger history, we need to take into account that low-mass galaxies are the most abundant type of galaxies at high-$z$. This means that an increase in the initial seeding probability will have a greater effect on them, making them much more likely to host a central BH. Hence, we can argue that the larger the number of low-mass galaxies seeded with central BHs, the larger the number of small galaxies undergoing a merger event where a BH is deposited, and therefore the smaller the average stellar mass of that type of galaxy. Also, a boosted seeding model is related to a highly active merger history that forces an earlier transition of central BHs into wanderers. In such cases, the galaxies hosting them would not have as much time to grow as those affected by a fiducial seeding. On the other hand, a weaker seeding is related to an early Universe characterized by a merger-deprived galactic evolution, where galaxies that contain central BHs have more time to merge with each other until they find one that indeed hosts a central BH and triggers a wandering event. 

\subsubsection{The population of leftover seeds in MW-type galaxies}
\label{seeding prescription leftover seeds}
We study the effects of varying the BH seeding prescription in the leftover seed population. In the boosted model, a typical MW galaxy hosts $\sim$80$\%$ more leftover seeds, with masses $\sim$5$\%$ smaller for the non-pairing seeds and $\sim$10$\%$ smaller for the pairing ones. In the weaker model, a typical MW galaxy hosts 50$\%$ fewer leftover seeds, with masses $\sim$10$\%$ smaller for the non-pairing seeds and $\sim$40$\%$ greater for the pairing ones. As per the locations of the leftover seeds within the galaxy, we find again invariability except for two of the WBH populations of the boosted model which are located farther from the galactic centre compared to the fiducial model; $\sim$35$\%$ in the case of the pairing population, and $\sim$45$\%$ in the case of the recoiled one. In terms of the ratio between the overall population of leftover seeds and non-leftover seeds, all three models are fairly consistent, showing that $\sim$60$\%$ of the total population has not grown more than twice its original mass. However, when we delve into the different types of WBHs, we find that in the disrupted and pairing populations, the ratio of leftover seeds to non-leftover seeds increases as the seeding probability of the model becomes higher. In contrast, this ratio decreases for recoiled WBHs when moving from the fiducial to the boosted model; but due to insufficient data on recoiled wanderers in the weaker model, we cannot draw a correlation. In any case, a higher ratio of leftover seeds in the boosted model implies a higher ratio of lower-mass WBHs that experience less dynamical friction (resulting in longer times to reach the galactic centre) and also that are prone to more powerful ejection kicks during a recoiled event. This could explain why disrupted and pairing leftover seeds appear farther away from the galactic centre as the seeding probability of the model increases.

\begin{figure*}
     \centering
     \begin{subfigure}[b]{0.33\textwidth}
         \centering
         \includegraphics[width=\textwidth]{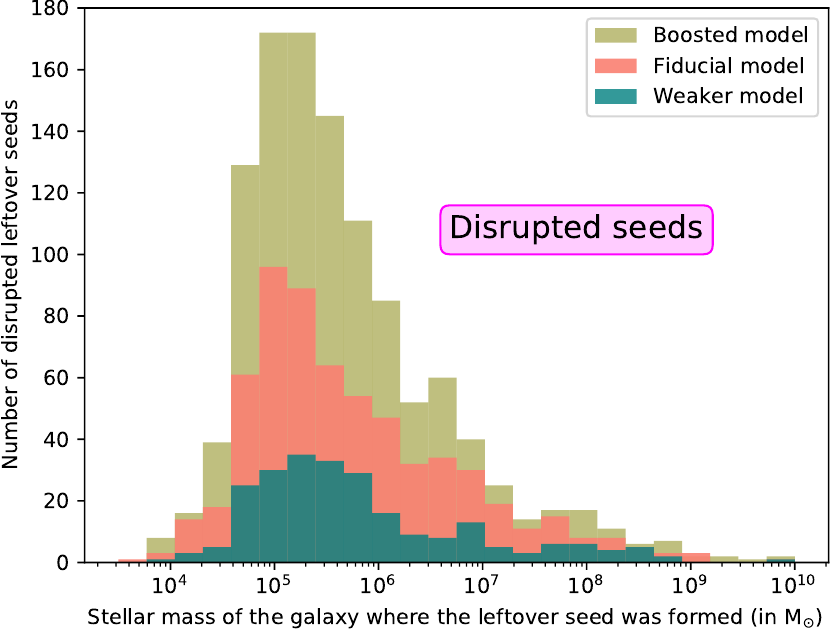}
     \end{subfigure}
     \hfill
     \begin{subfigure}[b]{0.33\textwidth}
         \centering
         \includegraphics[width=\textwidth]{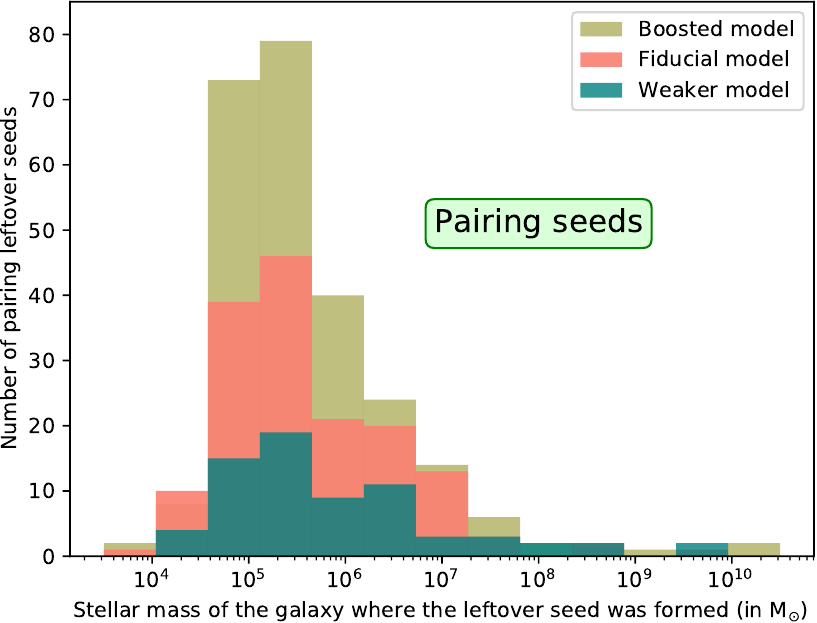}
     \end{subfigure}
     \hfill
     \begin{subfigure}[b]{0.33\textwidth}
         \centering
         \includegraphics[width=\textwidth]{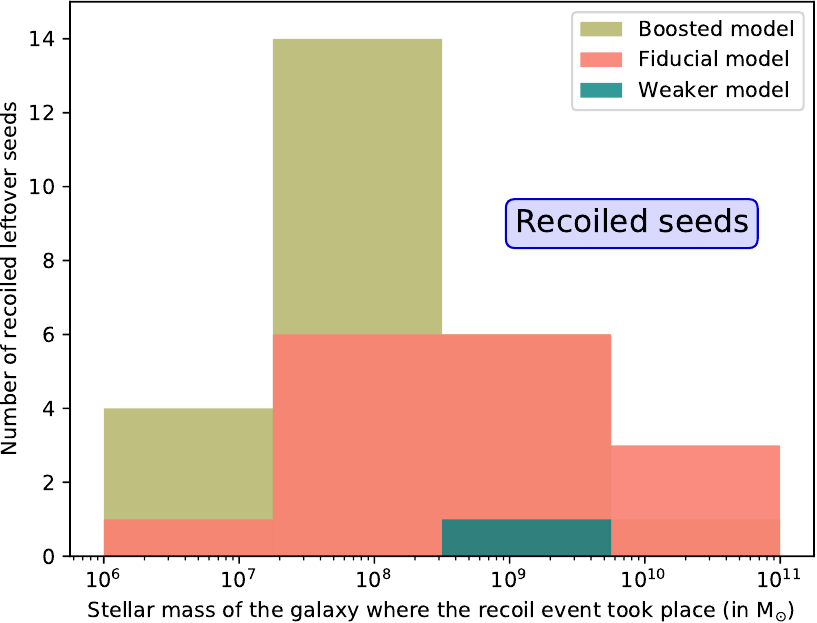}
     \end{subfigure}
        \caption{Number of disrupted (left), pairing (middle), and recoiled (right) leftover seeds as a function of the stellar mass of the galaxy that hosted them originally. In the case of the recoiled seeds (right), the x-axis does not represent the stellar mass of the galaxy where the leftover seeds were first formed, but rather the galaxy with which the BH-BH fusion and the subsequent recoil took place. The fiducial model is shown in red, while a version with boosted BH seeding is shown in olive and a weaker one in blue (see \S \ref{BH seeding} for more details).}
        \label{leftover seeds seeding prescription triple plot}. 
\end{figure*}

As per the effect of the seeding prescription in the leftover seeds' merger history, Fig. \ref{leftover seeds seeding prescription triple plot} displays the distribution of the number of leftover seeds per bin of stellar mass of the galaxy that hosted them when becoming wanderers. The same inverse correlation between seeding probability and stellar mass can be observed also for the leftover seeds. In the disrupted population, the stellar mass of the host galaxy increased from 2.8 $\times$ 10$^{5}$ M$_{\odot}$ in the boosted model, to 3.1 $\times$ 10$^{5}$ M$_{\odot}$ in the fiducial and 3.6 $\times$ 10$^{5}$ M$_{\odot}$ in the weaker one. The same applies for the pairing population, where the stellar mass increases from 2.3 $\times$ 10$^{5}$ M$_{\odot}$ in the boosted model, to 2.4 $\times$ 10$^{5}$ M$_{\odot}$ in the fiducial and 3.9 $\times$ 10$^{5}$ M$_{\odot}$ in the weaker one. Even though the recoiled population has too little data compared to the other two, it is clear that the inverse correlation also applies for the leftover seed population.

\section{Conclusions}
\label{Conclusions good}

We have utilized the semi-analytical model \texttt{L-Galaxies} to characterize the WBH population of MW-type galaxies at $z$ = 0. We divide the wandering population in three categories: disrupted WBHs (BHs from satellite galaxies that got disrupted by tidal forces upon entering a massive DM halo), pairing WBHs (BHs hosted by satellite galaxies that managed to reach and merge with a central galaxy and are now sinking towards the galactic core) and recoiled WBHs (BHs that got ejected into a bound orbit as the result of a pairing WBH merging with the central BH of a galaxy). We also study the population of leftover seeds, which in this work are defined as WBHs that have not grown more than twice their original seeding mass. We extract the following conclusions:

\begin{itemize}
    \item A typical MW galaxy is expected to host $\sim$10 WBHs, $\sim$70$\%$ of which have undergone a disruption event, $\sim$25$\%$ are in the process of forming a binary system with the central BH and $\sim$5$\%$ have undergone a recoil event. They account for $\sim$2$\%$ of the total BH mass budget of the galaxy.
    
    \item The locations of WBHs within the galaxy correlate with their formation scenario. While pairing WBHs are concentrated at $\lesssim$1 kpc from the galactic nucleus on the stellar disc, disrupted and recoiled are found $\gtrsim$100 kpc away from it. Such small and large distances might hinder their observational detection and explain the absence of strong observational evidence of WBHs in the MW.

    \item A typical WBH has an expected mass of $\sim$2 $\times$ 10$^{3}$ M$_{\odot}$, but recoiled WBHs, which have undergone a coalescence event, are predicted to have a mass one order of magnitude higher.

    \item While the total number of expected WBHs in a MW-type galaxy is in agreement with the prediction by \cite{tremmel2018} using the hydrodynamic simulation \texttt{ROMULUS25}, we find a higher spread in the radial distribution. This could be explained by \texttt{ROMULUS25} having higher seed masses and thus WBHs probably infalling in shorter time-scales. Also, \texttt{L-Galaxies} takes into account the recoil scenario when studying the WBH populations. 
    
    \item A typical MW galaxy is expected to host $\sim$6 leftover seed BHs, which we define in this work as WBHs that have not outgrown more than twice their original seeding mass. Non-pairing leftover seeds are located $\sim$14$\%$ farther from the galactic centre compared to their non-leftover counterparts. Pairing seeds, however, are expected $\sim$70$\%$ closer to the galactic centre. 

    \item Leftover seeds come from more metal-poor galaxies than non-leftover seeds (--1.4 compared to --0.8). This difference in the metallicity of the stellar environment surrounding them could be used when searching for hyper-compact clusters \citep{greene2021}.

    \item We have also analysed the impact of varying the BH seeding prescription of \texttt{L-Galaxies}. We find that a higher seeding prescription leads to an increase of low-mass galaxies hosting central BHs at high-$z$. This in turn provokes that more satellite galaxies are likely to leave their central BH as wanderers of the new DM host.
\end{itemize}

Overall we find that WBHs are either located within 1 kpc of the galactic centre ---most likely on the stellar disc--- or more than 100 kpc away from it, making them difficult to detect. The wide spread in metallicity found in the galaxies that hosted the WBHs before they became wanderers could be an indicator of the range of metallicities of the compact stellar populations that are predicted to be bound around WBHs. In a future work we will explore this in more depth by modeling clusters of stars around the WBHs.  

\section{Data Availability}
This work is based on an extension of the 2015 public version of the Munich model of galaxy formation and evolution \texttt{L-Galaxies}, available at http://galformod.mpagarching.mpg.de/public/LGalaxies/. 

\section{Acknowledgements}
S.B. acknowledges support from the Spanish Ministerio de Ciencia e Innovación through project PID2021-124243NB-C21. D.I.V. acknowledges the financial support provided under the European Union’s H2020 ERC Consolidator Grant ``Binary Massive Black Hole Astrophysics'' (B Massive, Grant Agreement: 818691). M.M. acknowledges support from the Spanish Ministry of Science and Innovation through the project PID2021-124243NB-C22. This project was also partially supported by the program Unidad de Excelencia María de Maeztu CEX2020-001058-M. D.S. acknowledges support from the National Key R$\&$D Program of China (grant No. 2018YFA0404503), the National Science Foundation of China (grant No. 12073014), the science research grants from the China Manned Space Project with No. CMS-CSST-2021-A05, and the Tsinghua University Initiative Scientific Research Program (No. 20223080023).  




\bibliographystyle{mnras}
\bibliography{references} 

\begin{thebibliography}{}
\makeatletter
\relax
\def\mn@urlcharsother{\let\do\@makeother \do\$\do\&\do\#\do\^\do\_\do\%\do\~}
\def\mn@doi{\begingroup\mn@urlcharsother \@ifnextchar [ {\mn@doi@}
  {\mn@doi@[]}}
\def\mn@doi@[#1]#2{\def\@tempa{#1}\ifx\@tempa\@empty \href
  {http://dx.doi.org/#2} {doi:#2}\else \href {http://dx.doi.org/#2} {#1}\fi
  \endgroup}
\def\mn@eprint#1#2{\mn@eprint@#1:#2::\@nil}
\def\mn@eprint@arXiv#1{\href {http://arxiv.org/abs/#1} {{\tt arXiv:#1}}}
\def\mn@eprint@dblp#1{\href {http://dblp.uni-trier.de/rec/bibtex/#1.xml}
  {dblp:#1}}
\def\mn@eprint@#1:#2:#3:#4\@nil{\def\@tempa {#1}\def\@tempb {#2}\def\@tempc
  {#3}\ifx \@tempc \@empty \let \@tempc \@tempb \let \@tempb \@tempa \fi \ifx
  \@tempb \@empty \def\@tempb {arXiv}\fi \@ifundefined
  {mn@eprint@\@tempb}{\@tempb:\@tempc}{\expandafter \expandafter \csname
  mn@eprint@\@tempb\endcsname \expandafter{\@tempc}}}

\bibitem[\protect\citeauthoryear{Adams et~al.,}{Adams et~al.}{2023}]{adams2023}
Adams N.,  et~al., 2023, Monthly Notices of the Royal Astronomical Society,
  518, 4755

\bibitem[\protect\citeauthoryear{Banados et~al.,}{Banados
  et~al.}{2018}]{bañados2018}
Banados E.,  et~al., 2018, Nature, 553, 473

\bibitem[\protect\citeauthoryear{Barrows, Mezcua  \& Comerford}{Barrows
  et~al.}{2019}]{barrows2019}
Barrows R.~S.,  Mezcua M.,   Comerford J.~M.,  2019, The Astrophysical Journal,
  882, 181

\bibitem[\protect\citeauthoryear{Bash, Gebhardt, Goss  \& Bout}{Bash
  et~al.}{2007}]{bash2007}
Bash F.,  Gebhardt K.,  Goss W.,   Bout P.~V.,  2007, The Astronomical Journal,
  135, 182

\bibitem[\protect\citeauthoryear{Baumgardt et~al.,}{Baumgardt
  et~al.}{2019}]{baumgardt2019}
Baumgardt H.,  et~al., 2019, Monthly Notices of the Royal Astronomical Society,
  488, 5340

\bibitem[\protect\citeauthoryear{Begelman, Blandford  \& Rees}{Begelman
  et~al.}{1980}]{begelman1980}
Begelman M.~C.,  Blandford R.~D.,   Rees M.~J.,  1980, Nature, 287, 307

\bibitem[\protect\citeauthoryear{Bellovary, Governato, Quinn, Wadsley, Shen  \&
  Volonteri}{Bellovary et~al.}{2010}]{bellovary2010}
Bellovary J.~M.,  Governato F.,  Quinn T.~R.,  Wadsley J.,  Shen S.,
  Volonteri M.,  2010, The Astrophysical Journal Letters, 721, L148

\bibitem[\protect\citeauthoryear{Binney \& Tremaine}{Binney \&
  Tremaine}{1987}]{binney1987}
Binney J.,  Tremaine S.,  1987, Galactic dynamics.
Princeton University Press

\bibitem[\protect\citeauthoryear{Blecha, Cox, Loeb  \& Hernquist}{Blecha
  et~al.}{2011}]{blecha2011}
Blecha L.,  Cox T.~J.,  Loeb A.,   Hernquist L.,  2011, Monthly Notices of the
  Royal Astronomical Society, 412, 2154

\bibitem[\protect\citeauthoryear{Bohn, Canalizo, Veilleux  \& Liu}{Bohn
  et~al.}{2021}]{bohn2021}
Bohn T.,  Canalizo G.,  Veilleux S.,   Liu W.,  2021, The Astrophysical
  Journal, 911, 70

\bibitem[\protect\citeauthoryear{Bonetti, Haardt, Sesana  \& Barausse}{Bonetti
  et~al.}{2018}]{bonetti2018}
Bonetti M.,  Haardt F.,  Sesana A.,   Barausse E.,  2018, Monthly Notices of
  the Royal Astronomical Society, 477, 3910

\bibitem[\protect\citeauthoryear{Bonoli, Marulli, Springel, White, Branchini
  \& Moscardini}{Bonoli et~al.}{2009}]{bonoli2009}
Bonoli S.,  Marulli F.,  Springel V.,  White S.~D.,  Branchini E.,   Moscardini
  L.,  2009, Monthly Notices of the Royal Astronomical Society, 396, 423

\bibitem[\protect\citeauthoryear{Boylan-Kolchin, Springel, White, Jenkins  \&
  Lemson}{Boylan-Kolchin et~al.}{2009}]{boylan2009}
Boylan-Kolchin M.,  Springel V.,  White S.~D.,  Jenkins A.,   Lemson G.,  2009,
  Monthly Notices of the Royal Astronomical Society, 398, 1150

\bibitem[\protect\citeauthoryear{Bromm \& Loeb}{Bromm \&
  Loeb}{2003}]{bromm2003}
Bromm V.,  Loeb A.,  2003, The Astrophysical Journal, 596, 34

\bibitem[\protect\citeauthoryear{Castellano et~al.,}{Castellano
  et~al.}{2022}]{castellano2022}
Castellano M.,  et~al., 2022, The Astrophysical Journal Letters, 938, L15

\bibitem[\protect\citeauthoryear{Chilingarian, Katkov, Zolotukhin, Grishin,
  Beletsky, Boutsia  \& Osip}{Chilingarian et~al.}{2018}]{chilingarian2018}
Chilingarian I.~V.,  Katkov I.~Y.,  Zolotukhin I.~Y.,  Grishin K.~A.,  Beletsky
  Y.,  Boutsia K.,   Osip D.~J.,  2018, The Astrophysical Journal, 863, 1

\bibitem[\protect\citeauthoryear{Coleman~Miller \& Hamilton}{Coleman~Miller \&
  Hamilton}{2002}]{coleman2002}
Coleman~Miller M.,  Hamilton D.~P.,  2002, Monthly Notices of the Royal
  Astronomical Society, 330, 232

\bibitem[\protect\citeauthoryear{Comerford, Pooley, Barrows, Greene, Zakamska,
  Madejski  \& Cooper}{Comerford et~al.}{2015}]{comerford2015}
Comerford J.~M.,  Pooley D.,  Barrows R.~S.,  Greene J.~E.,  Zakamska N.~L.,
  Madejski G.~M.,   Cooper M.~C.,  2015, The Astrophysical Journal, 806, 219

\bibitem[\protect\citeauthoryear{Croton et~al.,}{Croton
  et~al.}{2006}]{croton2006}
Croton D.~J.,  et~al., 2006, Monthly Notices of the Royal Astronomical Society,
  365, 11

\bibitem[\protect\citeauthoryear{Cseh, Kaaret, Corbel, K{\"o}rding, Coriat,
  Tzioumis  \& Lanzoni}{Cseh et~al.}{2010}]{cseh2010}
Cseh D.,  Kaaret P.,  Corbel S.,  K{\"o}rding E.,  Coriat M.,  Tzioumis A.,
  Lanzoni B.,  2010, Monthly Notices of the Royal Astronomical Society, 406,
  1049

\bibitem[\protect\citeauthoryear{Davis, Narayan, Zhu, Barret, Farrell, Godet,
  Servillat  \& Webb}{Davis et~al.}{2011}]{davis2011}
Davis S.~W.,  Narayan R.,  Zhu Y.,  Barret D.,  Farrell S.~A.,  Godet O.,
  Servillat M.,   Webb N.~A.,  2011, The Astrophysical Journal, 734, 111

\bibitem[\protect\citeauthoryear{Decarli et~al.,}{Decarli
  et~al.}{2018}]{decarli2018}
Decarli R.,  et~al., 2018, The Astrophysical Journal, 854, 97

\bibitem[\protect\citeauthoryear{Devecchi \& Volonteri}{Devecchi \&
  Volonteri}{2009}]{devecchi2009}
Devecchi B.,  Volonteri M.,  2009, The Astrophysical Journal, 694, 302

\bibitem[\protect\citeauthoryear{Dong et~al.,}{Dong et~al.}{2007}]{dong2007}
Dong X.,  et~al., 2007, The Astrophysical Journal, 657, 700

\bibitem[\protect\citeauthoryear{Dong, Ho, Yuan, Wang, Fan, Zhou  \&
  Jiang}{Dong et~al.}{2012}]{dong2012}
Dong X.-B.,  Ho L.~C.,  Yuan W.,  Wang T.-G.,  Fan X.,  Zhou H.,   Jiang N.,
  2012, The Astrophysical Journal, 755, 167

\bibitem[\protect\citeauthoryear{Eisenstein \& Loeb}{Eisenstein \&
  Loeb}{1994}]{eisenstein1994}
Eisenstein D.~J.,  Loeb A.,  1994, arXiv preprint astro-ph/9401016

\bibitem[\protect\citeauthoryear{Fan et~al.,}{Fan et~al.}{2001}]{fan2001}
Fan X.,  et~al., 2001, The Astronomical Journal, 122, 2833

\bibitem[\protect\citeauthoryear{Fan et~al.,}{Fan et~al.}{2003}]{fan2003}
Fan X.,  et~al., 2003, The Astronomical Journal, 125, 1649

\bibitem[\protect\citeauthoryear{Farrell, Webb, Barret, Godet  \&
  Rodrigues}{Farrell et~al.}{2009}]{farrell2009}
Farrell S.~A.,  Webb N.~A.,  Barret D.,  Godet O.,   Rodrigues J.~M.,  2009,
  Nature, 460, 73

\bibitem[\protect\citeauthoryear{Ferrara, Salvadori, Yue  \&
  Schleicher}{Ferrara et~al.}{2014}]{ferrara2014}
Ferrara A.,  Salvadori S.,  Yue B.,   Schleicher D.,  2014, Monthly Notices of
  the Royal Astronomical Society, 443, 2410

\bibitem[\protect\citeauthoryear{Filippenko \& Sargent}{Filippenko \&
  Sargent}{1989}]{filippenko1989}
Filippenko A.~V.,  Sargent W.~L.,  1989, Astrophysical Journal, Part 2-Letters
  (ISSN 0004-637X), vol. 342, July 1, 1989, p. L11-L14. Research supported by
  the University of California., 342, L11

\bibitem[\protect\citeauthoryear{Fischer, Iserlohe, Zuther, Bertram,
  Straubmeier, Sch{\"o}del  \& Eckart}{Fischer et~al.}{2006}]{fischer2006}
Fischer S.,  Iserlohe C.,  Zuther J.,  Bertram T.,  Straubmeier C.,
  Sch{\"o}del R.,   Eckart A.,  2006, Astronomy \& Astrophysics, 452, 827

\bibitem[\protect\citeauthoryear{Fryer, Woosley  \& Heger}{Fryer
  et~al.}{2001}]{fryer2001}
Fryer C.,  Woosley S.,   Heger A.,  2001, The Astrophysical Journal, 550, 372

\bibitem[\protect\citeauthoryear{Giallongo et~al.,}{Giallongo
  et~al.}{2015}]{giallongo2015}
Giallongo E.,  et~al., 2015, Astronomy \& Astrophysics, 578, A83

\bibitem[\protect\citeauthoryear{Giersz, Leigh, Hypki, L{\"u}tzgendorf  \&
  Askar}{Giersz et~al.}{2015}]{giersz2015}
Giersz M.,  Leigh N.,  Hypki A.,  L{\"u}tzgendorf N.,   Askar A.,  2015,
  Monthly Notices of the Royal Astronomical Society, 454, 3150

\bibitem[\protect\citeauthoryear{Gonz{\'a}lez \& Guzm{\'a}n}{Gonz{\'a}lez \&
  Guzm{\'a}n}{2018}]{gonzalez2018}
Gonz{\'a}lez J.,  Guzm{\'a}n F.,  2018, Physical Review D, 97, 063001

\bibitem[\protect\citeauthoryear{Greene}{Greene}{2012}]{greene2012}
Greene J.~E.,  2012, Nature Communications, 3, 1304

\bibitem[\protect\citeauthoryear{Greene \& Ho}{Greene \& Ho}{2004}]{greene2004}
Greene J.~E.,  Ho L.~C.,  2004, The Astrophysical Journal, 610, 722

\bibitem[\protect\citeauthoryear{Greene \& Ho}{Greene \& Ho}{2007}]{greene2007}
Greene J.~E.,  Ho L.~C.,  2007, The Astrophysical Journal, 670, 92

\bibitem[\protect\citeauthoryear{Greene et~al.,}{Greene
  et~al.}{2019}]{greene2019}
Greene J.~E.,  et~al., 2019, arXiv preprint arXiv:1903.08670

\bibitem[\protect\citeauthoryear{Greene, Strader  \& Ho}{Greene
  et~al.}{2020}]{greene2020}
Greene J.~E.,  Strader J.,   Ho L.~C.,  2020, Annual Review of Astronomy and
  Astrophysics, 58, 257

\bibitem[\protect\citeauthoryear{Greene et~al.,}{Greene
  et~al.}{2021}]{greene2021}
Greene J.~E.,  et~al., 2021, The Astrophysical Journal, 917, 17

\bibitem[\protect\citeauthoryear{Guo et~al.,}{Guo et~al.}{2011}]{guo2011}
Guo Q.,  et~al., 2011, Monthly Notices of the Royal Astronomical Society, 413,
  101

\bibitem[\protect\citeauthoryear{Haggard, Cool, Heinke, Van~der Marel, Cohn,
  Lugger  \& Anderson}{Haggard et~al.}{2013}]{haggard2013}
Haggard D.,  Cool A.~M.,  Heinke C.~O.,  Van~der Marel R.,  Cohn H.~N.,  Lugger
  P.~M.,   Anderson J.,  2013, The Astrophysical Journal Letters, 773, L31

\bibitem[\protect\citeauthoryear{Heger \& Woosley}{Heger \&
  Woosley}{2002}]{heger2002}
Heger A.,  Woosley S.~E.,  2002, The Astrophysical Journal, 567, 532

\bibitem[\protect\citeauthoryear{Heger, Fryer, Woosley, Langer  \&
  Hartmann}{Heger et~al.}{2003}]{heger2003}
Heger A.,  Fryer C.~L.,  Woosley S.~E.,  Langer N.,   Hartmann D.~H.,  2003,
  The Astrophysical Journal, 591, 288

\bibitem[\protect\citeauthoryear{Henriques, White, Thomas, Angulo, Guo, Lemson,
  Springel  \& Overzier}{Henriques et~al.}{2015}]{henriques2015}
Henriques B.~M.,  White S.~D.,  Thomas P.~A.,  Angulo R.,  Guo Q.,  Lemson G.,
  Springel V.,   Overzier R.,  2015, Monthly Notices of the Royal Astronomical
  Society, 451, 2663

\bibitem[\protect\citeauthoryear{{Hopkins} \& {Hernquist}}{{Hopkins} \&
  {Hernquist}}{2009}]{hopkins2009}
{Hopkins} P.~F.,  {Hernquist} L.,  2009, \mn@doi [\apj]
  {10.1088/0004-637X/698/2/1550}, \href
  {https://ui.adsabs.harvard.edu/abs/2009ApJ...698.1550H} {698, 1550}

\bibitem[\protect\citeauthoryear{Inayoshi, Visbal  \& Haiman}{Inayoshi
  et~al.}{2020}]{inayoshi2020}
Inayoshi K.,  Visbal E.,   Haiman Z.,  2020, Annual Review of Astronomy and
  Astrophysics, 58, 27

\bibitem[\protect\citeauthoryear{Izotov, Thuan  \& Guseva}{Izotov
  et~al.}{2007}]{izotov2007}
Izotov Y.~I.,  Thuan T.~X.,   Guseva N.~G.,  2007, The Astrophysical Journal,
  671, 1297

\bibitem[\protect\citeauthoryear{{Izquierdo-Villalba}, {Bonoli}, {Spinoso},
  {Rosas-Guevara}, {Henriques}  \&
  {Hern{\'a}ndez-Monteagudo}}{{Izquierdo-Villalba}
  et~al.}{2019}]{izquierdo2019}
{Izquierdo-Villalba} D.,  {Bonoli} S.,  {Spinoso} D.,  {Rosas-Guevara} Y.,
  {Henriques} B. M.~B.,   {Hern{\'a}ndez-Monteagudo} C.,  2019, \mn@doi
  [\mnras] {10.1093/mnras/stz1694}, \href
  {https://ui.adsabs.harvard.edu/abs/2019MNRAS.488..609I} {488, 609}

\bibitem[\protect\citeauthoryear{Izquierdo-Villalba, Bonoli, Dotti, Sesana,
  Rosas-Guevara  \& Spinoso}{Izquierdo-Villalba et~al.}{2020}]{izquierdo2020}
Izquierdo-Villalba D.,  Bonoli S.,  Dotti M.,  Sesana A.,  Rosas-Guevara Y.,
  Spinoso D.,  2020, Monthly Notices of the Royal Astronomical Society, 495,
  4681

\bibitem[\protect\citeauthoryear{{Izquierdo-Villalba}, {Sesana}, {Bonoli}  \&
  {Colpi}}{{Izquierdo-Villalba} et~al.}{2022}]{izquierdo2022}
{Izquierdo-Villalba} D.,  {Sesana} A.,  {Bonoli} S.,   {Colpi} M.,  2022,
  \mn@doi [\mnras] {10.1093/mnras/stab3239}, \href
  {https://ui.adsabs.harvard.edu/abs/2022MNRAS.509.3488I} {509, 3488}

\bibitem[\protect\citeauthoryear{{Izquierdo-Villalba}, {Colpi}, {Volonteri},
  {Spinoso}, {Bonoli}  \& {Sesana}}{{Izquierdo-Villalba}
  et~al.}{2023}]{izquierdo2023}
{Izquierdo-Villalba} D.,  {Colpi} M.,  {Volonteri} M.,  {Spinoso} D.,  {Bonoli}
  S.,   {Sesana} A.,  2023, \mn@doi [\aap] {10.1051/0004-6361/202347008}, \href
  {https://ui.adsabs.harvard.edu/abs/2023A&A...677A.123I} {677, A123}

\bibitem[\protect\citeauthoryear{Jiang, Dekel, Freundlich, van~den Bosch,
  Green, Hopkins, Benson  \& Du}{Jiang et~al.}{2021}]{jiang2021}
Jiang F.,  Dekel A.,  Freundlich J.,  van~den Bosch F.~C.,  Green S.~B.,
  Hopkins P.~F.,  Benson A.,   Du X.,  2021, Monthly Notices of the Royal
  Astronomical Society, 502, 621

\bibitem[\protect\citeauthoryear{Jonker, Torres, Fabian, Heida, Miniutti  \&
  Pooley}{Jonker et~al.}{2010}]{jonker2010}
Jonker P.,  Torres M.,  Fabian A.,  Heida M.,  Miniutti G.,   Pooley D.,  2010,
  arXiv preprint arXiv:1004.5379

\bibitem[\protect\citeauthoryear{Kaaret, Prestwich, Zezas, Murray, Kim,
  Kilgard, Schlegel  \& Ward}{Kaaret et~al.}{2001}]{kaaret2001}
Kaaret P.,  Prestwich A.,  Zezas A.,  Murray S.,  Kim D.-W.,  Kilgard R.,
  Schlegel E.,   Ward M.,  2001, Monthly Notices of the Royal Astronomical
  Society, 321, L29

\bibitem[\protect\citeauthoryear{Kim et~al.,}{Kim et~al.}{2015}]{kim2015}
Kim M.,  et~al., 2015, The Astrophysical Journal, 814, 8

\bibitem[\protect\citeauthoryear{King \& Dehnen}{King \&
  Dehnen}{2005}]{king2005}
King A.~R.,  Dehnen W.,  2005, Monthly Notices of the Royal Astronomical
  Society, 357, 275

\bibitem[\protect\citeauthoryear{K{\i}z{\i}ltan, Baumgardt  \&
  Loeb}{K{\i}z{\i}ltan et~al.}{2017}]{kiziltan2017}
K{\i}z{\i}ltan B.,  Baumgardt H.,   Loeb A.,  2017, Nature, 542, 203

\bibitem[\protect\citeauthoryear{Kong}{Kong}{2007}]{kong2007}
Kong A.~K.,  2007, The Astrophysical Journal, 661, 875

\bibitem[\protect\citeauthoryear{Kruijssen, Pfeffer, Reina-Campos, Crain  \&
  Bastian}{Kruijssen et~al.}{2019}]{kruijssen2019}
Kruijssen J.~D.,  Pfeffer J.~L.,  Reina-Campos M.,  Crain R.~A.,   Bastian N.,
  2019, Monthly Notices of the Royal Astronomical Society, 486, 3180

\bibitem[\protect\citeauthoryear{Kunth, Sargent  \& Bothun}{Kunth
  et~al.}{1987}]{kunth1987}
Kunth D.,  Sargent W.,   Bothun G.,  1987, Astronomical Journal (ISSN
  0004-6256), vol. 93, Jan. 1987, p. 29-32., 93, 29

\bibitem[\protect\citeauthoryear{Licquia \& Newman}{Licquia \&
  Newman}{2013}]{licquia2013}
Licquia T.,  Newman J.,  2013, in American Astronomical Society Meeting
  Abstracts\# 221. pp 254--11

\bibitem[\protect\citeauthoryear{Lin et~al.,}{Lin et~al.}{2016}]{lin2016}
Lin D.,  et~al., 2016, The Astrophysical Journal, 821, 25

\bibitem[\protect\citeauthoryear{Lin et~al.,}{Lin et~al.}{2018}]{lin2018}
Lin D.,  et~al., 2018, Nature Astronomy, 2, 656

\bibitem[\protect\citeauthoryear{Lin et~al.,}{Lin et~al.}{2020}]{lin2020}
Lin D.,  et~al., 2020, The Astrophysical Journal Letters, 892, L25

\bibitem[\protect\citeauthoryear{Lodato \& Natarajan}{Lodato \&
  Natarajan}{2006}]{lodato2006}
Lodato G.,  Natarajan P.,  2006, Monthly Notices of the Royal Astronomical
  Society, 371, 1813

\bibitem[\protect\citeauthoryear{Loeb \& Rasio}{Loeb \& Rasio}{1994}]{loeb1994}
Loeb A.,  Rasio F.~A.,  1994, arXiv preprint astro-ph/9401026

\bibitem[\protect\citeauthoryear{{Lousto}, {Zlochower}, {Dotti}  \&
  {Volonteri}}{{Lousto} et~al.}{2012}]{Lousto2012}
{Lousto} C.~O.,  {Zlochower} Y.,  {Dotti} M.,   {Volonteri} M.,  2012, \mn@doi
  [\prd] {10.1103/PhysRevD.85.084015}, \href
  {https://ui.adsabs.harvard.edu/abs/2012PhRvD..85h4015L} {85, 084015}

\bibitem[\protect\citeauthoryear{Maccarone \& Servillat}{Maccarone \&
  Servillat}{2008}]{maccarone2008}
Maccarone T.~J.,  Servillat M.,  2008, Monthly Notices of the Royal
  Astronomical Society, 389, 379

\bibitem[\protect\citeauthoryear{Maccarone, Fender  \& Tzioumis}{Maccarone
  et~al.}{2005}]{maccarone2005}
Maccarone T.~J.,  Fender R.~P.,   Tzioumis A.~K.,  2005, Monthly Notices of the
  Royal Astronomical Society: Letters, 356, L17

\bibitem[\protect\citeauthoryear{Madau \& Rees}{Madau \&
  Rees}{2001}]{madau2001}
Madau P.,  Rees M.~J.,  2001, The Astrophysical Journal, 551, L27

\bibitem[\protect\citeauthoryear{Maiolino et~al.,}{Maiolino
  et~al.}{2024}]{maiolino2024}
Maiolino R.,  et~al., 2024, Nature, 627, 59

\bibitem[\protect\citeauthoryear{Mapelli}{Mapelli}{2016}]{mapelli2016}
Mapelli M.,  2016, Monthly Notices of The Royal Astronomical Society, 459, 3432

\bibitem[\protect\citeauthoryear{Marulli, Bonoli, Branchini, Gilli, Moscardini
  \& Springel}{Marulli et~al.}{2009}]{marulli2009}
Marulli F.,  Bonoli S.,  Branchini E.,  Gilli R.,  Moscardini L.,   Springel
  V.,  2009, Monthly Notices of the Royal Astronomical Society, 396, 1404

\bibitem[\protect\citeauthoryear{Matsumoto, Tsuru, Koyama, Awaki, Canizares,
  Kawai, Matsushita  \& Kawabe}{Matsumoto et~al.}{2001}]{matsumoto2001}
Matsumoto H.,  Tsuru T.,  Koyama K.,  Awaki H.,  Canizares C.,  Kawai N.,
  Matsushita S.,   Kawabe R.,  2001, The Astrophysical Journal, 547, L25

\bibitem[\protect\citeauthoryear{Matsuoka et~al.,}{Matsuoka
  et~al.}{2019}]{matsuoka2019}
Matsuoka Y.,  et~al., 2019, The Astrophysical Journal Letters, 872, L2

\bibitem[\protect\citeauthoryear{Mayer \& Bonoli}{Mayer \&
  Bonoli}{2018}]{mayer2018}
Mayer L.,  Bonoli S.,  2018, Reports on Progress in Physics, 82, 016901

\bibitem[\protect\citeauthoryear{Mezcua}{Mezcua}{2017}]{mezcua2017}
Mezcua M.,  2017, International Journal of Modern Physics D, 26, 11

\bibitem[\protect\citeauthoryear{Mezcua}{Mezcua}{2019}]{mezcua2019}
Mezcua M.,  2019, Nature Astronomy, 3, 6

\bibitem[\protect\citeauthoryear{Mezcua}{Mezcua}{2020}]{mezcua2020feedback}
Mezcua M.,  2020, Proceedings of the International Astronomical Union, 15, 238

\bibitem[\protect\citeauthoryear{Mezcua \& Lobanov}{Mezcua \&
  Lobanov}{2011}]{mezcua2011}
Mezcua M.,  Lobanov A.,  2011, Astronomische Nachrichten, 332, 379

\bibitem[\protect\citeauthoryear{Mezcua \& S{\'a}nchez}{Mezcua \&
  S{\'a}nchez}{2020}]{mezcua2020broad}
Mezcua M.,  S{\'a}nchez H.~D.,  2020, The Astrophysical Journal Letters, 898,
  L30

\bibitem[\protect\citeauthoryear{Mezcua \& S{\'a}nchez}{Mezcua \&
  S{\'a}nchez}{2024}]{mezcua2024}
Mezcua M.,  S{\'a}nchez H.~D.,  2024, Monthly Notices of the Royal Astronomical
  Society, p. stae292

\bibitem[\protect\citeauthoryear{Mezcua, Farrell, Gladstone  \& Lobanov}{Mezcua
  et~al.}{2013a}]{mezcua2013a}
Mezcua M.,  Farrell S.,  Gladstone J.~C.,   Lobanov A.,  2013a, Monthly Notices
  of the Royal Astronomical Society, 436, 1546

\bibitem[\protect\citeauthoryear{Mezcua, Roberts, Sutton  \& Lobanov}{Mezcua
  et~al.}{2013b}]{mezcua2013b}
Mezcua M.,  Roberts T.,  Sutton A.,   Lobanov A.,  2013b, Monthly Notices of
  the Royal Astronomical Society, 436, 3128

\bibitem[\protect\citeauthoryear{Mezcua, Roberts, Lobanov  \& Sutton}{Mezcua
  et~al.}{2015}]{mezcua2015}
Mezcua M.,  Roberts T.,  Lobanov A.,   Sutton A.~D.,  2015, Monthly Notices of
  the Royal Astronomical Society, 448, 1893

\bibitem[\protect\citeauthoryear{Mezcua, Civano, Fabbiano, Miyaji  \&
  Marchesi}{Mezcua et~al.}{2016}]{mezcua2016}
Mezcua M.,  Civano F.,  Fabbiano G.,  Miyaji T.,   Marchesi S.,  2016, The
  Astrophysical Journal, 817, 20

\bibitem[\protect\citeauthoryear{Mezcua, Kim, Ho  \& Lonsdale}{Mezcua
  et~al.}{2018}]{mezcua2018}
Mezcua M.,  Kim M.,  Ho L.,   Lonsdale C.,  2018, Monthly Notices of the Royal
  Astronomical Society: Letters, 480, L74

\bibitem[\protect\citeauthoryear{Mezcua, Suh  \& Civano}{Mezcua
  et~al.}{2019}]{mezcua2019z3.4}
Mezcua M.,  Suh H.,   Civano F.,  2019, Monthly Notices of the Royal
  Astronomical Society, 488, 685

\bibitem[\protect\citeauthoryear{Mortlock et~al.,}{Mortlock
  et~al.}{2011}]{mortlock2011}
Mortlock D.~J.,  et~al., 2011, Nature, 474, 616

\bibitem[\protect\citeauthoryear{Murphy, Yates  \& Mohamed}{Murphy
  et~al.}{2022}]{murphy2022}
Murphy G.~G.,  Yates R.~M.,   Mohamed S.~S.,  2022, Monthly Notices of the
  Royal Astronomical Society, 510, 1945

\bibitem[\protect\citeauthoryear{Naidu et~al.,}{Naidu et~al.}{2022}]{naidu2022}
Naidu R.~P.,  et~al., 2022, The Astrophysical Journal Letters, 940, L14

\bibitem[\protect\citeauthoryear{Nguyen et~al.,}{Nguyen
  et~al.}{2018}]{nguyen2018}
Nguyen D.~D.,  et~al., 2018, The Astrophysical Journal, 858, 118

\bibitem[\protect\citeauthoryear{O'Leary \& Loeb}{O'Leary \&
  Loeb}{2009}]{oleary2009}
O'Leary R.~M.,  Loeb A.,  2009, Monthly Notices of the Royal Astronomical
  Society, 395, 781

\bibitem[\protect\citeauthoryear{O'Leary \& Loeb}{O'Leary \&
  Loeb}{2012}]{oleary2012}
O'Leary R.~M.,  Loeb A.,  2012, Monthly Notices of the Royal Astronomical
  Society, 421, 2737

\bibitem[\protect\citeauthoryear{O’leary, Rasio, Fregeau, Ivanova  \&
  O’Shaughnessy}{O’leary et~al.}{2006}]{oleary2006}
O’leary R.~M.,  Rasio F.~A.,  Fregeau J.~M.,  Ivanova N.,   O’Shaughnessy
  R.,  2006, The Astrophysical Journal, 637, 937

\bibitem[\protect\citeauthoryear{Pacucci, Seepaul, Ni, Cappelluti  \&
  Foord}{Pacucci et~al.}{2024}]{pacucci2024}
Pacucci F.,  Seepaul B.,  Ni Y.,  Cappelluti N.,   Foord A.,  2024, Universe,
  10, 225

\bibitem[\protect\citeauthoryear{Pasham, Strohmayer  \& Mushotzky}{Pasham
  et~al.}{2014}]{pasham2014}
Pasham D.~R.,  Strohmayer T.~E.,   Mushotzky R.~F.,  2014, Nature, 513, 74

\bibitem[\protect\citeauthoryear{Planck~Collaboration
  et~al.,}{Planck~Collaboration et~al.}{2014}]{planckcollaboration2014}
Planck~Collaboration P.,  et~al., 2014, A\&A, 571, A16

\bibitem[\protect\citeauthoryear{Pooley \& Rappaport}{Pooley \&
  Rappaport}{2006}]{pooley2006}
Pooley D.,  Rappaport S.,  2006, The Astrophysical Journal Letters, 644, L45

\bibitem[\protect\citeauthoryear{Rashkov \& Madau}{Rashkov \&
  Madau}{2013}]{rashkov2013}
Rashkov V.,  Madau P.,  2013, The Astrophysical Journal, 780, 187

\bibitem[\protect\citeauthoryear{Reines \& Comastri}{Reines \&
  Comastri}{2016}]{reines2016}
Reines A.~E.,  Comastri A.,  2016, Publications of the Astronomical Society of
  Australia, 33, e054

\bibitem[\protect\citeauthoryear{Reines, Sivakoff, Johnson  \& Brogan}{Reines
  et~al.}{2011}]{reines2011}
Reines A.~E.,  Sivakoff G.~R.,  Johnson K.~E.,   Brogan C.~L.,  2011, Nature,
  470, 66

\bibitem[\protect\citeauthoryear{Reines, Greene  \& Geha}{Reines
  et~al.}{2013}]{reines2013}
Reines A.~E.,  Greene J.~E.,   Geha M.,  2013, The Astrophysical Journal, 775,
  116

\bibitem[\protect\citeauthoryear{Reines, Plotkin, Russell, Mezcua, Condon,
  Sivakoff  \& Johnson}{Reines et~al.}{2014}]{reines2014}
Reines A.~E.,  Plotkin R.~M.,  Russell T.~D.,  Mezcua M.,  Condon J.~J.,
  Sivakoff G.~R.,   Johnson K.~E.,  2014, The Astrophysical Journal Letters,
  787, L30

\bibitem[\protect\citeauthoryear{Reines, Condon, Darling  \& Greene}{Reines
  et~al.}{2020}]{reines2020}
Reines A.~E.,  Condon J.~J.,  Darling J.,   Greene J.~E.,  2020, The
  Astrophysical Journal, 888, 36

\bibitem[\protect\citeauthoryear{Ricarte, Tremmel, Natarajan, Zimmer  \&
  Quinn}{Ricarte et~al.}{2021a}]{ricarte2021}
Ricarte A.,  Tremmel M.,  Natarajan P.,  Zimmer C.,   Quinn T.,  2021a, Monthly
  Notices of the Royal Astronomical Society, 503, 6098

\bibitem[\protect\citeauthoryear{Ricarte, Tremmel, Natarajan  \& Quinn}{Ricarte
  et~al.}{2021b}]{ricarte2021b}
Ricarte A.,  Tremmel M.,  Natarajan P.,   Quinn T.,  2021b, The Astrophysical
  Journal Letters, 916, L18

\bibitem[\protect\citeauthoryear{Salehirad, Reines  \& Molina}{Salehirad
  et~al.}{2022}]{salehirad2022}
Salehirad S.,  Reines A.~E.,   Molina M.,  2022, The Astrophysical Journal,
  937, 7

\bibitem[\protect\citeauthoryear{Seepaul, Pacucci  \& Narayan}{Seepaul
  et~al.}{2022}]{seepaul2022}
Seepaul B.~S.,  Pacucci F.,   Narayan R.,  2022, Monthly Notices of the Royal
  Astronomical Society, 515, 2110

\bibitem[\protect\citeauthoryear{Silk \& Arons}{Silk \& Arons}{1975}]{silk1975}
Silk J.,  Arons J.,  1975, The Astrophysical Journal, 200, L131

\bibitem[\protect\citeauthoryear{Soria, Hau  \& Pakull}{Soria
  et~al.}{2013}]{soria2013}
Soria R.,  Hau G.~K.,   Pakull M.~W.,  2013, The Astrophysical Journal Letters,
  768, L22

\bibitem[\protect\citeauthoryear{Spera \& Mapelli}{Spera \&
  Mapelli}{2017}]{spera2017}
Spera M.,  Mapelli M.,  2017, Monthly Notices of the Royal Astronomical
  Society, 470, 4739

\bibitem[\protect\citeauthoryear{Spinoso, Bonoli, Valiante, Schneider  \&
  Izquierdo-Villalba}{Spinoso et~al.}{2023}]{spinoso2023}
Spinoso D.,  Bonoli S.,  Valiante R.,  Schneider R.,   Izquierdo-Villalba D.,
  2023, Monthly Notices of the Royal Astronomical Society, 518, 4672

\bibitem[\protect\citeauthoryear{Springel, White, Tormen  \&
  Kauffmann}{Springel et~al.}{2001}]{springel2001}
Springel V.,  White S.~D.,  Tormen G.,   Kauffmann G.,  2001, Monthly Notices
  of the Royal Astronomical Society, 328, 726

\bibitem[\protect\citeauthoryear{Springel et~al.,}{Springel
  et~al.}{2005}]{springel2005}
Springel V.,  et~al., 2005, nature, 435, 629

\bibitem[\protect\citeauthoryear{Strader, Chomiuk, Maccarone, Miller-Jones,
  Seth, Heinke  \& Sivakoff}{Strader et~al.}{2012}]{strader2012}
Strader J.,  Chomiuk L.,  Maccarone T.~J.,  Miller-Jones J.~C.,  Seth A.~C.,
  Heinke C.~O.,   Sivakoff G.~R.,  2012, The Astrophysical Journal Letters,
  750, L27

\bibitem[\protect\citeauthoryear{Sutherland \& Dopita}{Sutherland \&
  Dopita}{1993}]{sutherland1993}
Sutherland R.~S.,  Dopita M.~A.,  1993, The Astrophysical Journal Supplement
  Series, 88, 253

\bibitem[\protect\citeauthoryear{Sutton, Roberts, Walton, Gladstone  \&
  Scott}{Sutton et~al.}{2012}]{sutton2012}
Sutton A.~D.,  Roberts T.~P.,  Walton D.~J.,  Gladstone J.~C.,   Scott A.~E.,
  2012, Monthly Notices of the Royal Astronomical Society, 423, 1154

\bibitem[\protect\citeauthoryear{Toki \& Takada}{Toki \&
  Takada}{2021}]{toki2021}
Toki S.,  Takada M.,  2021, arXiv preprint arXiv:2103.13015

\bibitem[\protect\citeauthoryear{Tremmel, Governato, Volonteri, Pontzen  \&
  Quinn}{Tremmel et~al.}{2018}]{tremmel2018}
Tremmel M.,  Governato F.,  Volonteri M.,  Pontzen A.,   Quinn T.~R.,  2018,
  The Astrophysical Journal Letters, 857, L22

\bibitem[\protect\citeauthoryear{Tremou et~al.,}{Tremou
  et~al.}{2018}]{tremou2018}
Tremou E.,  et~al., 2018, The Astrophysical Journal, 862, 16

\bibitem[\protect\citeauthoryear{Ulvestad, Greene  \& Ho}{Ulvestad
  et~al.}{2007}]{ulvestad2007}
Ulvestad J.~S.,  Greene J.~E.,   Ho L.~C.,  2007, The Astrophysical Journal
  Letters, 661, L151

\bibitem[\protect\citeauthoryear{Venemans et~al.,}{Venemans
  et~al.}{2013}]{venemans2013}
Venemans B.,  et~al., 2013, The Astrophysical Journal, 779, 24

\bibitem[\protect\citeauthoryear{Volonteri}{Volonteri}{2010}]{volonteri2010}
Volonteri M.,  2010, The Astronomy and Astrophysics Review, 18, 279

\bibitem[\protect\citeauthoryear{Volonteri \& Perna}{Volonteri \&
  Perna}{2005}]{volonteri2005}
Volonteri M.,  Perna R.,  2005, Monthly Notices of the Royal Astronomical
  Society, 358, 913

\bibitem[\protect\citeauthoryear{Volonteri, Haardt  \& Madau}{Volonteri
  et~al.}{2003}]{volonteri2003}
Volonteri M.,  Haardt F.,   Madau P.,  2003, The Astrophysical Journal, 582,
  559

\bibitem[\protect\citeauthoryear{Volonteri, Habouzit  \& Colpi}{Volonteri
  et~al.}{2021}]{volonteri2021}
Volonteri M.,  Habouzit M.,   Colpi M.,  2021, Nature Reviews Physics, 3, 732

\bibitem[\protect\citeauthoryear{Wang et~al.,}{Wang et~al.}{2021}]{wang2021}
Wang F.,  et~al., 2021, The Astrophysical Journal Letters, 907, L1

\bibitem[\protect\citeauthoryear{Webb et~al.,}{Webb et~al.}{2012}]{webb2012}
Webb N.,  et~al., 2012, Science, 337, 554

\bibitem[\protect\citeauthoryear{Weller, Pacucci, Hernquist  \& Bose}{Weller
  et~al.}{2022}]{weller2022}
Weller E.~J.,  Pacucci F.,  Hernquist L.,   Bose S.,  2022, Monthly Notices of
  the Royal Astronomical Society, 511, 2229

\bibitem[\protect\citeauthoryear{White \& Frenk}{White \&
  Frenk}{1991}]{white1991}
White S.~D.,  Frenk C.~S.,  1991, Astrophysical Journal, Part 1 (ISSN
  0004-637X), vol. 379, Sept. 20, 1991, p. 52-79. Research supported by NASA,
  NSF, and SERC., 379, 52

\bibitem[\protect\citeauthoryear{White \& Rees}{White \&
  Rees}{1978}]{white1978}
White S.~D.,  Rees M.~J.,  1978, Monthly Notices of the Royal Astronomical
  Society, 183, 341

\bibitem[\protect\citeauthoryear{Willott et~al.,}{Willott
  et~al.}{2007}]{willott2007}
Willott C.~J.,  et~al., 2007, The Astronomical Journal, 134, 2435

\bibitem[\protect\citeauthoryear{Willott et~al.,}{Willott
  et~al.}{2010}]{willott2010}
Willott C.~J.,  et~al., 2010, The Astronomical Journal, 139, 906

\bibitem[\protect\citeauthoryear{Wrobel \& Nyland}{Wrobel \&
  Nyland}{2020}]{wrobel2020}
Wrobel J.,  Nyland K.,  2020, arXiv preprint arXiv:2007.12093

\bibitem[\protect\citeauthoryear{Wrobel, Nyland  \& Miller-Jones}{Wrobel
  et~al.}{2015}]{wrobel2015}
Wrobel J.,  Nyland K.,   Miller-Jones J.,  2015, The Astronomical Journal, 150,
  120

\bibitem[\protect\citeauthoryear{Wu et~al.,}{Wu et~al.}{2015}]{wu2015}
Wu X.-B.,  et~al., 2015, Nature, 518, 512

\makeatother
\end{thebibliography}





\label{lastpage}
\end{document}